\documentclass[a4paper,12pt]{article}

\input cohmacros.sty

\def\at#1{[*** \att #1 ***]}
\def\at#1{} 
\usepackage{citeref}

\advance\textheight2.2cm        
\advance\topmargin-2.0cm
\advance\textwidth2.6cm
\advance\evensidemargin-0.3cm
\advance\oddsidemargin-0.3cm

\begin{document}

\vspace*{-2cm}
\begin{center}
{\LARGE \bf Introduction to coherent spaces} \\

\vspace{1cm}

\centerline{\sl {\large \bf Arnold Neumaier}}

\vspace{0.5cm}

\centerline{\sl Fakult\"at f\"ur Mathematik, Universit\"at Wien}
\centerline{\sl Oskar-Morgenstern-Platz 1, A-1090 Wien, Austria}
\centerline{\sl email: Arnold.Neumaier@univie.ac.at}
\centerline{\sl WWW: \url{http://www.mat.univie.ac.at/~neum}}

\end{center}

\vspace{0.5cm}

\hfill arXiv:1804.01402

\hfill September 28, 2018

\vspace{0.5cm}

\bigskip
\bfi{Abstract.}
The notion of a coherent space is a nonlinear version of the notion of
a complex Euclidean space: The vector space axioms are dropped while
the notion of inner product is kept.

Coherent spaces provide a setting for the study of geometry in a
different direction than traditional metric, topological, and
differential geometry. Just as it pays to study the properties of
manifolds independently of their embedding into a Euclidean space,
so it appears fruitful to study the properties of coherent spaces
independent of their embedding into a Hilbert space.

Coherent spaces have close relations to reproducing kernel Hilbert
spaces, Fock spaces, and unitary group representations, and to many
other fields of mathematics, statistics, and physics.

This paper is the first of a series of papers and defines concepts and
basic theorems about coherent spaces, associated vector spaces, and
their topology. Later papers in the series discuss symmetries of
coherent spaces, relations to homogeneous spaces, the theory of group
representations, $C^*$-algebras, hypergroups, finite geometry,
and applications to quantum physics. While the applications to quantum
physics were the main motiviation for developing the theory, many more
applications exist in complex analysis, group theory, probability
theory, statistics, physics, and engineering.

For the discussion of questions concerning coherent spaces, please use
the discussion forum \\
\url{https://www.physicsoverflow.org}.


\newpage
\tableofcontents 

\newpage
\section{Introduction}\label{s.intro}

In this paper, the first one of a series of papers on coherent spaces,
the length and angular properties of vectors in a Euclidean space
(embodied in the inner product) are generalized in a similar way as,
in the past, metric properties of Euclidean spaces were generalized to
metric spaces, differential properties of Euclidean spaces were
generalized to manifolds, and topological properties of Euclidean spaces
were generalized to topological spaces.

The notion of a coherent space is a nonlinear version of the notion of
a complex Euclidean space: The vector space axioms are dropped while
the inner product -- now called a coherent product -- is kept.
Thus coherent spaces provide a setting for the study of geometry in a
different direction than traditional metric, topological, and
differential geometry. Just as it pays to study the properties of
manifolds independently of their embedding into a Euclidean space,
so it appears fruitful to study the properties of coherent spaces
independent of their embedding into a Hilbert space.

However, coherent spaces (especially in the particular form of coherent
manifolds) may also be viewed as a new, geometric way of working with
concrete Hilbert spaces in which they are embeddable.
In place of measures and integration dominating traditional techniques
based on Hilbert spaces of functions, differentiation turns out to be
the basic tool for evaluating inner products and matrix elements of
linear operators.

One of the strengths of the coherent space approach is that it makes
many different things look alike. Coherent spaces have close relations
to many important fields of mathematics, statistics, physics, and
engineering. A theory of coherent spaces will provide a unified
geometric view of these applications.
Coherent spaces combine the rich, often highly characteristic variety of
symmetries of traditional geometric structures with the computational
tractability of traditional tools from numerical analysis and
statistics.

Coherent spaces give a natural geometric setting to the concept of
coherent states. In particular, the compact symmetric spaces (and many
noncompact ones) appear naturally as coherent spaces when equipped with
a coherent product derived from the coherent states on semisimple Lie
groups (cf. \sca{Perelomov} \cite{Per}). In these cases, the coherent
product is naturally related to the differential, metric, symplectic,
and K\"ahler structure of the associated symmetric spaces
(cf. \sca{Zhang} et al. \cite[Sections IIIC1 and VI]{ZhaFG}, and
Subsection \ref{ss.BWset} below).

Certain coherent spaces are closely related to quantum field theory
(\sca{Baez} et al. \cite{BaeSZ}, \sca{Glimm \& Jaffe} \cite{GliJ})
and the theory of Hida distributions in the white noise calculus
for classical stochastic processes (\sca{Hida \& Si} \cite{HidS},
\sca{Hida \& Streit} \cite{HidSt}, \sca{Obata} \cite{Oba}).

As we shall see in this paper, coherent spaces abstract the essential
geometric properties needed to define a reproducing kernel Hilbert
space. Examples of reproducing kernels (i.e., what in the present
context are coherent products) were first discussed by \sca{Zaremba}
\cite{Zar} in the context of boundary value problems and by \sca{Mercer}
\cite{Mer} in the context of integral equations. The theory was
systematically developed by \sca{Aronszajn} \cite{Aro0,Aro},
\sca{Krein} \cite{Kre1,Kre2}, and others. For a history see \sca{Berg}
et al. \cite{BerCR} and \sca{Stewart} \cite{Ste}.

Coherent spaces and reproducing kernel Hilbert spaces are mathematically
almost equivalent concepts, and there is a vast literature related to
the latter. Most relevant for the present sequence of papers are the
books by \sca{Perelomov} \cite{Per}, \sca{Neeb} \cite{Nee}, and
\sca{Neretin} \cite{Ner}.
However, the emphasis in these books is quite different from the
present exposition, as they are primarily interested in properties of
the associated function spaces and group representations, while we are
primarily interested in the geometry and symmetry properties and in
computational tractability.

Of particular importance is the use of reproducing kernels in complex
analysis (see, e.g., \sca{Faraut \& Kor\'anyi}, \cite{FarK},
\sca{Upmeier} \cite{Upm}, and \sca{de Branges} \cite{deBra})
and group theory (see, e.g., \sca{Neeb} \cite{Nee}),
where they are the basis of many important theorems.

One of the important uses of coherent spaces is that many Euclidean
spaces are described most simply and naturally in terms of a nice,
small subset of coherent states, and all their properties can be
investigated in terms of the associated coherent space.
Already \sca{Glauber} \cite[p.2771]{Gla}, who coined the notion of a
coherent state, mentioned that ''The scalar product may, in fact, be
calculated more simply than by using wave functions'', and the same
can be said for almost everything one wants to calculate in the
applications of coherent states.

In particular, while the study of most problems in traditional function
spaces for applications rely heavily on measures and integration,
the quantum spaces of coherent spaces with an easily computable
coherent product can be studied efficiently without measures or
integrations, in terms of the explicit coherent product and
differentiation only; cf. Subsection \ref{ss.states} below.
This makes many calculations easy that are difficult in Hilbert spaces
whose inner product is defined through a measure.
This will be substantiated in \sca{Neumaier \& Farashahi}
\cite{NeuF.cohQQ,NeuF.cohMan} and other papers of this series.

Coherent states are most often discussed as being parameterized by
points on a connected manifold. But the concept of a
coherent space also makes sense in a nontrivial way for finite spaces.
There are strong relations between finite coherent spaces, finite
metric spaces, graphs, and combinatorial designs.
See \sca{Bekka \& de la Harpe} \cite{BekH},
\sca{Brouwer} et al. \cite{BroCN}, \sca{Godsil} \cite{God},
\sca{Neumaier} \cite{Neu.dist,Neu.graphReg,Neu.conf,Neu.derEV}.
This shows that the concept of coherent spaces provides a nontrivial
extension of the theory of coherent states, in this respect similar to
that of the measure-free coherent states of \sca{Horzela \& Szafraniec}
\cite{HorS}.

\bigskip
Reproducing kernel Hilbert spaces and the associated coherent states
also have applications in many other fields of
mathematics (see, e.g., \cite{AglYY,Aka,AliAG,Alp,AlpDR,AntAL,BerCR,
ComR,KlaS,ParS,Per,Sai1,Sai2}),
statistics and stochastic processes (see, e.g., \cite{BerT,HidS,HidSt,
Oba,ParS,Rad,Sai2}),
physics (see, e.g., \cite{AglYY,AlpDR,AntAL,ComR,Gaz,InoKG,KlaS,Per}),
and engineering (see, e.g., \cite{AliAG,Gaz}).

In particular, there are relations to\\
(i) Christoffel--Darboux kernels for orthogonal polynomials,\\
(ii) Euclidean representations of finite geometries,\\
(iii) zonal spherical functions on symmetric spaces,\\
(iv) coherent states for Lie groups acting on homogeneous spaces,\\
(v) unitary representations of groups,\\
(vi) abstract harmonic analysis,\\
(vii) states of $C^*$-algebras in functional analysis,\\
(viii) reproducing kernel Hilbert spaces in complex analysis,\\
(ix) Pick--Nevanlinna interpolation theory,\\
(x) transfer functions in control theory,\\
(xi) positive definite kernels for radial basis functions,\\
(xii) positive definite kernels in data mining,\\
(xiii) positive definite functions in probability theory,\\
(xvi) exponential families in probability theory and statistics,\\
(xv) the theory of random matrices,\\
(xvi) Hida distributions for white noise analysis,\\
(xvii) K\"ahler manifolds and geometric quantization,\\
(xviii) coherent states in quantum mechanics,\\
(xix) squeezed states in quantum optics,\\
(xx) inverse scattering in quantum mechanics,\\
(xxi) Hartree--Fock equations in quantum chemistry,\\
(xxii) mean field calculations in statistical mechanics,\\
(xxiii) path integrals in quantum mechanics,\\
(xxiv) functional integrals in quantum field theory,\\
(xxv) integrable quantum systems.

These relations will be established in a series of papers of which the
present one is the first, laying the foundations. The web site
\cite{cohSpaces.html} will display at any time the most recent state of
affairs. Other, in some draft form already existing, papers of this
series will discuss

\pt
the quantization of coherent spaces with a compatible manifold (or
stratification) structure (\sca{Neumaier \& Ghaani Farashahi}
\cite{NeuF.cohQuant,NeuF.cohQQ,NeuF.cohMan}) and related quantum
dynamics,

\pt
relations between coherent spaces and $C^*$-algebras
(\sca{Neumaier \& Ghaani Farashahi} \cite{NeuF.cohCstar}),

\pt
finite coherent spaces related to finite geometries (\sca{Neumaier}
\cite{Neu.cohFin}) and hypergroups (\sca{Neumaier \& Ghaani Farashahi}
\cite{NeuF.cohHyp}),and

\pt
the classical limit in coherent spaces (\sca{Neumaier}
\cite{Neu.cohSpec}) and related semiclassical expansions.

\bigskip
In this paper we only treat the most basic aspects of coherent spaces,
starting from first principles.
We define the concept, give some basic examples, and define the
corresponding quantum spaces.
We then discuss functions of positive type and the construction of the
quantum space, and give a long list of constructions of such functions,
useful for constructing new coherent spaces from old ones.
More advanced topics and applications will be given in later papers of
the series. Due to the introductory character of this paper all proofs
are carried out in detail.

To illustrate some of the connections to physics and complex analysis,
we give in Subsection \ref{ss.ex} a long but still very incomplete list
of examples of coherent spaces. Some of these examples are very
elementary and can be understood informally before reading the
systematic exposition of the theory. Many more coherent spaces can be
constructed by modifying given ones using the recipes from Subsection
\ref{ss.Constructions}.

\bigskip
One of the most important concepts for coherent spaces is that of their
symmetries. Indeed, most of the applications of coherent spaces in
quantum mechanics and quantum field theory rely on the presence of a
large symmetry group. In \sca{Neumaier \& Ghaani Farashahi}
\cite{NeuF.cohQuant}, the second paper of this series, we define the
notion of a coherent map of a coherent space, which will be exploited
in depth in later papers of this series. Invertible coherent maps form
the symmetry group of the coherent space, while all coherent maps often
form a bigger semigroup. In many concrete examples, these are related
to so-called Olshanski semigroups (\sca{Ol'shanskii} \cite{Ols}).
It is shown that one obtains a vast generalization of the theory of
normally ordered operator expressions in Fock spaces.

In \sca{Neumaier \& Ghaani Farashahi} \cite{NeuF.cohQQ,NeuF.cohMan}
we define coherent manifolds and coherent vector fields, the
infinitesimal analogue of coherent maps, and their quantization.
In \sca{Neumaier \& Ghaani Farashahi} \cite{NeuF.cohInv}, coherent
spaces with an additional compatible involution structure are
introduced and shown to lead to a natural generalization of the
geometric quantization of K\"ahler manifolds.

Note that some papers by \sca{Vourdas} \cite{Vou1,Vou2} use the term 
''coherent space'' for a different concept, also related to coherent 
states.
Completely unrelated is the notion of coherent spaces used in logic.

\bigskip
{\bf Acknowledgments.} Thanks to Arash Ghaani Farashahi,
Waltraud Huyer, Rahel Kn\"opfel, David Bar Moshe, Mike Mowbray,
Karl-Hermann Neeb, Hermann Schichl and Eric Wofsey for useful
discussions related to the subject.

\section{Euclidean spaces}\label{s.coh}

There is a notational discrepancy in how mathematicians and physicists
treat Hilbert spaces. In physics, one often works with
finite-dimensional Hilbert spaces treated as $\Cz^n$ and hence wants
to write the Hermitian inner product as $\<x,y\>=x^*y$. This definition
dictates the use of a Hermitian inner product that is antilinear in the
first argument, the convention followed, e.g., in
\sca{Reed \& Simon} \cite{ReeS.1}.
This is also the choice adopted in Dirac's bra-ket
notation whose usage in quantum mechanics is very widespread.

The notation in this paper was chosen to extend the traditional notation
of standard finite-dimensional matrix algebra as closely as possible to
arbitrary complex inner product spaces and associated linear operators.
In matrix algebra, column vectors and the corresponding matrices with
one column are identical objects, row vectors are the linear
functionals, and the adjoint is the conjugate transpose.
For example, $\Hz=\Hz^\*=\Cz^n$ is the space of column vectors of size
$n$, the dual space $\Hz^*$ is the space of row vectors of size $n$,
and the operator product $\phi^*\psi$ of a row vector $\phi^*$ and a
column vector $\psi$ is the standard Hermitian inner product of the
column vectors $\phi$ and $\psi$.
We use Greek lower case letters to write vectors, thus emphasizing
their intended use as quantum state vectors in quantum mechanics.

On the other hand, mathematicians working on reproducing kernel Hilbert
spaces use an inner product $(x,y)$ antilinear in the second argument,
related to the physicist's  inner product  $\<x,y\>$ by
$(x,y)=\ol{\<x,y\>}$. This is the convention followed, e.g., in
\sca{Rudin} \cite{Rud.FA}.
Although the two ways of defining the inner product lead to fully
equivalent theories, all details look a bit different, a fact that has
to be taken into account when reading the literature on the subject.
For example, in the description based on the physical tradition it is
preferable to work with the antidual space in place of the dual space
used in the mathematical tradition.

In functional analysis, linear operators in Hilbert spaces are usually
considered each with their own domain. But many computations in quantum
mechanics require the consideration of algebras of operators with a
common domain. The latter is a Euclidean space, a dense subspace of a
Hilbert space. This space and its antidual play in many respects a more
basic role in quantum physics than the Hilbert space itself.

Therefore, and to avoid possible confusion caused by the different
traditions we give in the present section a self-contained
introduction to Euclidean spaces and their associated spaces.
Moreover, in the final Section \ref{s.cohfuncpostype}, we give
statements of the relevant main results from the vast literature on
functions of positive type, rewritten in the present notation and with
detailed proofs.

\subsection{Euclidean spaces and their antidual}\label{ss.E}

A \bfi{Euclidean space} is a complex vector space $\Hz$ with a binary
operation that assigns to $\phi,\psi\in\Hz$ the
\bfi{Hermitian inner product} $\<\phi,\psi\>\in\Cz$, antilinear in the
first and linear in the second argument, such that
\lbeq{e.her}
\ol{\<\phi,\psi\>}=\<\psi,\phi\>,
\eeq
\lbeq{e.def}
\<\psi,\psi\>>0 \Forall \psi\in\Hz\setminus\{0\}.
\eeq
Here $\alpha>0$ says that the complex number $\alpha$ is real and
positive.

Since every Euclidean space can be completed to a Hilbert space
(cf. Theorem \ref{t.Hilbert} below), the Euclidean spaces are in fact
just the subspaces of Hilbert spaces, with the induced inner product.
However, it is of interest to develop the theory of Euclidean spaces
independently since some additional topological structure is
present that has no simple counterpart in the Hilbert space setting.

We define the \bfi{antidual} $\Hz^\*$ of $\Hz$ to be the vector space
of antilinear functionals $\phi:\Hz\to\Cz$.
We turn $\Hz^\*$ into a locally convex space (cf. \sca{Rudin}
\cite[Chapter 3]{Rud.FA}) with the \bfi{weak-* topology} induced by
the family of seminorms $|\cdot|_\psi$ with $\psi\in\Hz$ defined by
$|\phi|_\psi:=|\phi(\psi)|$ for $\phi\in\Hz^\*$.
Thus $U\subseteq \Hz^\*$ is a neighborhood of $\phi\in\Hz^\*$ iff
there are finitely many $\psi_k\in\Hz$ such that $U$ contains all
$\phi'\in\Hz^\*$ with $|\phi'(\psi_k)-\phi(\psi_k)|\le 1$ for all $k$.
(The $1$ can be replaced by any positive constant since the
$\psi_k$ can be arbitrarily scaled.) As a consequence, a net\footnote{
All limits are formulated in terms of nets indexed by a directed set
rather than sequences indexed by nonnegative integers, to cover the
possibility of nonseparable spaces.
In a separable Hilbert space, net convergence and sequence convergence
are equivalent. In general, there is a difference and nets are
needed to obtain the correct topology.
\\
For those not familiar with nets -- they are generalizations of
sequences defining the appropriate form of the limit in the
nonseparable case.
In the separable case, nets can always be replaced by sequences.
Thus readers will grasp the main content if, on first reading, they
simply think of nets as being sequences.
} 
of vectors $\phi_\ell\in\Hz^\*$ \bfi{converges} in the weak-* topology
to the \bfi{weak-* limit} $\phi\in\Hz^\*$ iff $
\phi_\ell(\psi)\to\phi(\psi)$ for all $\psi\in\Hz$. Because of
\gzit{e.def}, we may identify $\psi\in\Hz$ with the antilinear
functional on $\Hz$ defined by
\lbeq{e.psi}
\psi(\phi):=\<\phi,\psi\> \for \phi\in\Hz.
\eeq
This definition turns $\Hz$ canonically into a subspace of $\Hz^\*$.

\begin{expl}\label{ex.fs}
The vector space $M(Z)$ of complex-valued functions $\psi:Z\to\Cz$ with
finite support is with the inner product
\[
\<\phi,\psi\>:=\D\sum_{z\in Z}\ol{\phi(z)}\psi(z)
\]
a Euclidean space. The antidual $M(Z)^\*$ is the space of all
complex-valued functions $\psi:Z\to\Cz$, with
\[
\psi(\phi):=\D\sum_{z\in Z}\ol{\phi(z)}\psi(z).
\]
Weak-* convergence in $M(Z)$ is just pointwise convergence.
\end{expl}

\begin{prop}\label{p.HHstar}~\\
(i)  Every $\psi\in\Hz^\*$ is the weak-* limit of a net of vectors
from $\Hz$.

(ii) For every weak-* continuous antilinear functional $\Psi$ on a
subspace $V$ of $\Hz^\*$, 

there is a $\psi\in\Hz$ such that
\[
\Psi(\phi)=\ol{\phi(\psi)} \for \phi\in V.
\]
(iii) Every weak-* continuous antilinear functional on $\Hz$ has a
unique extension to a weak-* continuous antilinear functional on
$\Hz^\*$.
\end{prop}

\bepf
(i) For any finite-dimensional subspace $V$ of $\Hz$ there is a unique
$\psi_V\in V$ such that $\psi(\phi)=\<\phi,\psi_V\>$ for all
$\phi\in V$.
The collection of finite-dimensional subspaces form a directed set under
inclusion, hence the $\psi_V$ form a net. The net converges to $\psi$
in the weak-* topology since for all $\phi\in\Hz$,
\[
(\psi-\psi_V)(\phi)=\<\phi,\psi-\psi_V\>
=\<\phi,\psi\>-\<\phi,\psi_V\>\to 0.
\]
(ii) By continuity, there is a neighborhood $N$ of zero such that
\[
|\Psi(\phi)|\le 1 \Forall \phi \in N.
\]
By definition of the weak-* topology, there are
$\psi_1,\ldots,\psi_n\in\Hz$ such that $N$ contains all $\phi\in\Hz$
with $|\phi(\psi_k)|\le 1$ for $k=1,\ldots,n$. Let $A:V\to \Cz^n$ be
the linear mapping with $(A\phi)_k:=\phi(\psi_k)$ for all $k$.
If $A\phi=0$ and $\eps>0$ then $\eps^{-1}\phi\in N$, hence 
$|\Psi(\eps^{-1}\phi)|\le 1$. Thus $|\Psi(\phi)|\le \eps$ for all 
$\eps>0$, giving $\Psi(\phi)=0$. 

This implies
that $f(A\phi):=\Psi(\phi)$ defines an antilinear functional $f$ on the
range of $A$. We may extend $f$ to an antilinear functional on $\Cz^n$.
This has the form $f(x)=u^T\ol x$ with suitable $u\in\Cz^n$.
Now $\psi:=\sum u_k\psi_k$ is in $\Hz$ and satisfies
\[
\phi(\psi)=\phi\Big(\sum u_k\psi_k\Big)=\sum \ol u_k\phi(\psi_k)
=\sum \ol u_k(A\phi)_k,
\]
\[
\ol{\phi(\psi)}=\sum  u_k\ol{(A\phi)_k}=u^T\ol{A\phi}=f(A\phi)
=\Psi(\phi).
\]
(iii) This follows from (ii) for $V=\Hz$.
\epf

We define the \bfi{adjoint} $\psi^*$ of $\psi\in\Hz$ to be the
linear functional on $\Hz^\*$ that maps $\phi\in\Hz^\*$ to
\[
\psi^*\phi:=\phi(\psi),
\]
and the \bfi{adjoint} $\psi^*$ of $\psi\in\Hz^\*$ to be the
linear functional on $\Hz$ that maps $\phi\in\Hz$ to
\[
\psi^*\phi:=\ol{\psi(\phi)}.
\]
As a consequence,
\[
\<\phi,\psi\>=\phi^*\psi \for \phi,\psi\in\Hz.
\]
Moreover, if $\phi,\psi\in\Hz^\*$ and one of them is in $\Hz$ then
\[
\ol{\psi^*\phi}=\phi^*\psi.
\]

\begin{cor}\label{c.Hdense} ~\\
(i) Every linear mapping $f:\Hz\to\Cz$ can be written
in the form $f=\phi^*$ for some $\phi\in\Hz^\*$.

(ii) Every weak-* continuous linear functional $f$ on $\Hz^\*$ can be
written in the form $f=\phi^*$ for some $\phi\in\Hz$.
\end{cor}

\bepf
(i) The mapping $\phi:\Hz\to\Cz$ defined by $\phi(\psi):=\ol{f\psi}$
is antilinear, hence $\phi\in\Hz^\*$. Since
$f\psi=\ol{\phi(\psi)}=\phi^*\psi$ we conclude that
$f=\phi^*$.

(ii) follows in the same way from Proposition \ref{p.HHstar}(ii).
\epf

We equip $\Hz$ with the \bfi{strict topology}, the locally
convex topology in which all antilinear (and hence all linear)
functionals are continuous. Thus ${ }^*$ is an antiisomorphism from
$\Hz^\*$ to the space of all linear functionals on $\Hz$, the
\bfi{dual} of $\Hz$ with respect to the strict topology.

\begin{expl}\label{ex.fs2}
In the Euclidean space $M(Z)$ defined in Example \ref{ex.fs}, a net
$\psi_\ell$ in $M(Z)$ converges to $\psi\in M(Z)$ in the strict topology
iff $\psi$ is the weak-* limit and there is a finite subset $S$ of $Z$
such that $\psi_\ell(z)=\psi(z)$ for all $z\in Z\setminus S$.
\end{expl}

\subsection{Norm and completion of a Euclidean space}

\begin{prop}\label{p.norm}
The \bfi{Euclidean norm} $\|\psi\|$ defined on a Euclidean space $\Hz$
by
\[
\|\psi\|:=\sqrt{\psi^*\psi}
\]
is positive when $\psi\ne 0$. It satisfies for $\phi,\psi\in\Hz$ the
\bfi{Cauchy--Schwarz inequality}
\lbeq{e.CauchySchwarz}
|\phi^*\psi|\le \|\phi\|\,\|\psi\|,
\eeq
and the \bfi{triangle inequality}
\lbeq{e.triangle}
\|\phi+\psi\|\le \|\phi\|+\|\psi\|,
\eeq
and for $\lambda\in\Cz$ the relation
\[
\|\lambda\psi\|=|\lambda|\,\|\psi\|.
\]
\end{prop}

\bepf
\gzit{e.CauchySchwarz} holds for $\psi=0$. For $\psi\ne0$ and
$\beta:=\psi^*\phi/\|\psi\|^2$,
\[
0\le (\phi-\beta\psi)^*(\phi-\beta\psi)
=\|\phi\|^2-2\re(\beta\phi^*\psi)+|\beta|^2\|\psi\|^2
=\|\phi\|^2-|\phi^*\psi|^2/\|\psi\|^2,
\]
so that \gzit{e.CauchySchwarz} holds also in this case.
The triangle inequality now follows from
\[
\|\phi+\psi\|^2=\|\phi\|^2+2\re\phi^*\psi+\|\psi\|^2
\le \|\phi\|^2+2\|\phi\|\,\|\psi\|+\|\psi\|^2 =(\|\phi\|+\|\psi\|)^2.
\]
The final equation is obvious.
\epf

A mapping $f:\Hz\to\Cz$ is called \bfi{bounded} if there is a constant
$C$ such that $|f(\psi)|\le C\|\psi\|$ for all $\psi\in\Hz$.
A \bfi{Cauchy net} in $\Hz$ consists of a net $\psi_\ell$ in $\Hz$ such
that for every $\eps>0$ there is an index $N$ such that
$\|\psi_j-\psi_k\|\le \eps$ for $j,k\ge N$. It is called \bfi{bounded}
if $\D\sup_\ell \|\psi_\ell\|<\infty$.
A \bfi{Hilbert space} is a Euclidean space containing with each bounded
Cauchy net its weak-* limit.
\at{check \sca{Reed \& Simon} \cite{ReeS.1}}

\begin{thm}\label{t.Hilbert}
The set $\ol\Hz$ of all bounded antilinear functionals on $\Hz$
is a Hilbert space, and we have
\lbeq{e.HzHzx}
\Hz\subseteq \ol\Hz\subseteq \Hz^\*.
\eeq
If a net $\psi_\ell$ in $\ol\Hz$ has a weak-* limit $\psi\in\ol\Hz$
then $\|\psi_\ell-\psi\|\to 0$.
\end{thm}

\bepf
Clearly $\ol\Hz$ is a subspace of $\Hz^\*$. The Cauchy--Schwarz
inequality says that, as an antilinear functional, $\psi\in\Hz$ is
bounded. Thus $\ol\Hz$ contains $\Hz$, and \gzit{e.HzHzx} holds.

Now suppose that
$\phi,\psi\in\ol \Hz$ and that $\phi=\lim \phi_j$,
$\psi=\lim \psi_j$ for nets with $\phi_j,\psi_j\in\Hz$.
Then
\[
|\phi_j^*\psi_j-\phi_k^*\psi_k|
=\Big|(\phi_j-\phi_k)^*\psi_j+\phi_k^*(\psi_j-\psi_k)\Big|
\le \|\phi_j-\phi_k\|\,\|\psi_j\|+\|\phi_k\|\,\|\psi_j-\psi_k\|
\]
converges to zero as $j,k\to\infty$. Therefore the $\phi_j^*\psi_j$
form a Cauchy net and the limit
\lbeq{e.star}
\phi^*\psi:=\lim_{\ell} \<\phi_\ell,\psi_\ell\>
\eeq
exists.
A similar \at{}
argument shows that the limit is independent of the choice
of the nets. We take \gzit{e.star} as the definition of the inner
product in $\ol\Hz$. It is easy to see that the inner product is
Hermitian and linear in the second argument. \at{add details}
Thus $\ol\Hz$ is a Euclidean space. In particular, Proposition
\ref{p.norm} applies with $\ol\Hz$ in place of $\Hz$.

To show completeness, let $\phi_\ell$ be a bounded Cauchy net in
$\ol\Hz$. Then for every $\psi\in\Hz$,
\[
|\phi_\ell^*\psi-\phi_k^*\psi|=|(\phi_\ell-\phi_k)^*\psi|
\le \|\phi_\ell-\phi_k\|\,\|\psi\| \to 0 \for k,\ell\to\infty,
\]
hence the $\phi_\ell^*\psi$ form a Cauchy net in $\Cz$ and
converge. Thus
\[
f(\psi):=\lim_{\ell\to\infty} \phi_\ell^*\psi
\]
defines a map $f:\Hz\to\Cz$. Since for $\mu,\mu'\in\Cz$ and
$\psi,\psi'\in\Hz$,
\[
f(\mu\psi+\mu'\psi')-\mu f(\psi)-\mu'f(\psi')
=\lim_{\ell\to\infty} \Big(\phi_\ell^*(\mu\psi+\mu'\psi')
-\mu \phi_\ell^*\psi-\mu'\phi_\ell^*\psi'\Big)=0,
\]
$f$ is linear, and by Corollary \ref{c.Hdense}(i), $f=\phi^*$ for
some $\phi\in\Hz^\*$. Clearly $\phi$ is the weak-* limit of the
$\phi_\ell$. Since the Cauchy net is bounded, $\phi$ is bounded, too,
\at{add details}
hence $\phi\in\ol\Hz$.

To prove the final statement, we note that
\[
\|\psi_\ell-\psi\|^2=(\psi_\ell-\psi)^*(\psi_\ell-\psi_m)
   +(\psi_\ell-\psi)^*(\psi_m-\psi)
\le\|\psi_\ell-\psi\|\,\|\psi_\ell-\psi_m\|
   +(\psi_\ell-\psi)^*(\psi_m-\psi).
\]
The first term goes to zero due to the Cauchy property, and the second
term due to weak-* convergence to zero.
\epf

\begin{cor}\label{c.Riesz} (\bfi{Riesz representation theorem})\\
For every norm-continuous linear functional $f$ on $\ol\Hz$ there is a
vector $\psi\in\ol\Hz$ such that
\lbeq{e.Psiol}
f(\phi)=\psi^*\phi \Forall \phi\in\ol{\Hz}.
\eeq
\end{cor}

\bepf
The mapping $\psi:\ol\Hz\to\Cz$ defined by $\psi(\phi):=\ol{f(\phi)}$
for $\phi\in\Hz$ is antilinear, hence belongs to $\ol\Hz$.
\epf

We call $\ol\Hz$ the \bfi{completion} of $\Hz$.
If $\Hz$ is finite-dimensional then $\Hz=\ol\Hz=\Hz^\*$ by standard
arguments, and all topologies considered are equivalent.
If $\Hz$ is infinite-dimensional then we usually\footnote{
except when $\Hz$ is already an infinite-dimensional Hilbert space,
in which case $\Hz=\ol\Hz\ne \Hz^\*$
} 
have $\Hz\ne \ol\Hz\ne \Hz^\*$
For example, the space $\Hz:=C([-1,1])$ of continuous functions on
$[-1,1]$ has as antilinear functionals not only all elements
of the Hilbert space $\ol\Hz=L^2([-1,1])$ of square integrable
functions on $[-1,1]$ but also all function evaluation maps,
corresponding to distributions. All these are elements of the
antidual $\Hz^\*$.
In infinite dimensions, the norm topology in $\ol\Hz$ is weaker than
the strict topology in $\Hz$ but stronger than the weak-* topology in
$\Hz^\*$.

By now, $\phi^*\psi$ is defined whenever $\phi,\psi\in\Hz^\*$ and
either one of the two is in $\Hz$ or both are in $\ol\Hz$. Thus we have
a partial binary operation $\,{}^*\,$ on $\Hz^\*$, called the
\bfi{partial inner product} (\bfi{PIP}). It satisfies
\lbeq{e.her*}
\ol{\phi^*\psi}=\psi^*\phi.
\eeq
Unless $\Hz$ is
finite-dimensional, the partial inner product is not everywhere defined.
(For example, in the antidual of $C([-1,1])$, the inner product of two
delta distriutions at the same point is not defined.)
Proposition \ref{p.HHstar}(i) implies that $\Hz$ is dense in $\Hz^\*$.
In particular, $\Hz^\*$ is a positive definite \bfi{PIP space} in the
sense of \sca{Antoine \& Trapani} \cite{PIP}.

\begin{expl}\label{ex.fs3}
The completion $\ol{M(Z)}$ of the Euclidean space $M(Z)$ discussed in
Examples \ref{ex.fs} and \ref{ex.fs2} is the Hilbert space of functions
$\psi:Z\to\Cz$ with countable support and finite
$\D\sum_{z\in Z}|\psi(z)|^2$.
The inner product of $\phi,\psi\in\ol{M(Z)}$ is given by the absolutely
convergent countable sum
$\phi^*\psi:=\D\sum_{z\in Z}\ol{\phi(z)}\psi(z)$.
\end{expl}

\subsection{Linear mappings between Euclidean spaces}

An \bfi{isometry} from a Euclidean space $U$ to a Euclidean space $V$
is a linear map $A:U\to V$ such that
\[
\|A\psi\|=\|\psi\| \Forall \psi\in U.
\]
Isometries are injective since $A\psi=0$ implies $\|\psi\|=0$ and hence
$\psi=0$. An \bfi{isomorphism} from $U$ to $V$ is a surjective isometry
$A:U\to V$. Its inverse is an isomorphism from $V$ to $U$. if such an
isomorphism exists the Euclidean spaces $U$ and $V$ are called
\bfi{isometric} or \bfi{isomorphic}.

If $U$ and $V$ are (complex) topological vector spaces we write
$\Lin(U,V)$ for the vector space of all continuous linear mappings from
$U$ to $V$, and $\Lin U$ for $\Lin(U,U)$. We identify $V$ with the
space $\Lin(\Cz,V)$ via
\[
\psi\alpha:=\alpha \psi \for \alpha\in\Cz,~\psi\in V.
\]

\begin{prop}\label{A.ss=A}
Let $U$ and $V$ be Euclidean spaces.

(i) For any linear map $A:U\to V^\*$, the mapping $A^*\phi:U\to\Cz$
defined for $\phi\in V$ by
\[
(A^*\phi)(\psi):=(A\psi)^*\phi \for \psi\in U
\]
is an antilinear functional and defines an operator $A^*:V\to U^\*$
with
\[
(A\psi)^*\phi = \psi^*(A^*\phi),
\]
called the \bfi{adjoint} of $A$.

(ii) Any linear map $A:U\to V^\*$ is continuous, i.e.,
$A\in\Lin(U,V^\*)$.

(iii) The mapping $\,{}^*\,$ that maps $A$ to $A^*$ is an antilinear
mapping from $\Lin(U,V^\*)$ to $\Lin(V,U^\*)$ and satisfies
\[
A^{**}=A.
\]
\end{prop}

\bepf
(i) is obvious.

(ii) We need to show that for every weak-* neighborhood $N$ of $0$ in
$V^\*$, there is a strict neighborhood $M$ of $0$ in $U$ such that
$A\psi\in N$ for all $\psi\in M$.
By definition of the weak-* topology, there are
$\phi_1,\ldots,\phi_n\in V$ such that $N$ contains all $\phi\in V^\*$
with $|\phi(\phi_k)|\le 1$ for $k=1,\ldots,n$. The set $M$ of all
$\psi\in U$ with $|\psi(A^*\phi_k)|\le 1$ for $k=1,\ldots,n$ is a
strict neighborhood of 0 in $U$ and has the required property.

(iii) The dependence of $A^*\phi$ on $\phi$ is linear, thus the adjoint
$A^*$ is a linear operator. By (i), $A^*\in\Lin(V,U^\*)$.
Corollary \ref{c.Hdense}(ii) gives $V^{\*\*}=V$ and
$U^{\*\*}=U$. \at{check and simplify}
Hence $A^*:V\to U^\*$ is given by $A^*\psi(\phi)=(A\phi)^*\psi$ for
all $\psi\in V$ and $\phi\in U$. Thus we have for all $\phi\in U$ and
$\psi\in V$,
\[
A^{**}\phi(\psi)=(A^*\psi)^*\phi=A\phi(\psi),
\]
which implies that $A^{**}=A$.
\epf

Since $V\subseteq V^\*$, the adjoint is also defined for
$A\in \Lin(U,V)$ and then makes sense as a mapping
$A^*\in\Lin(V^\*,U^\*)$, and we have
\[
A^*B^*=(BA)^* \iif A\in \Lin(U,V),~B\in\Lin(V,W^\*).
\]
We write
\[
\Linx\Hz:=\Lin(\Hz,\Hz^\*)
\]
for the vector space of continuous linear operators from a Euclidean
space $\Hz$ to its antidual. Since $\Hz^{\*\*}=\Hz$ by Proposition 
\ref{p.HHstar}(v), we conclude:

\begin{cor}
If $A\in\Linx\Hz$ then $A^*\in\Linx\Hz$ and we have
\lbeq{e.sesqui}
\phi^*A\psi=(\phi^*A)\psi=\phi^*(A\psi)=(A^*\phi)^*\psi
\for \phi,\psi\in\Hz.
\eeq
Thus $\phi^*A\psi$ defines a sesquilinear form on $\Hz$.
\end{cor}

Here $\phi^*$ is treated as the adjoint $\phi^*:\Hz^\*\to \Cz$
of $\phi:\Cz\to\Hz$ under the identification $V=\Lin(\Cz,V)$.
We call $A\in\Linx\Hz$ \bfi{Hermitian} if $A^*=A$; then
$\ol{\phi^*A\psi}=\psi^*A\phi$, so that the associated sesquilinear
form is Hermitian.

\subsection{Functions of positive type}\label{ss.funcpostype}

A complex $n\times n$ matrix $G$ is \bfi{Hermitian} if
$\ol G_{jk}=G_{kj}$ for $j,k=1,\dots,n$, \bfi{positive semidefinite} if
$u^*Gu\ge 0$ for all $u\in\Cz^n$, and \bfi{conditionally semidefinite}
if $u^*Gu\ge 0$ for all $u\in\Cz^n$ with $\D\sum_k u_k=0$.

Let $Z$ be a nonempty set. We call a function $F:Z\times Z\to \Cz$
\bfi{of positive type} (resp. \bfi{conditionally positive}) over $Z$
if, for every finite sequence $z_1,\dots,z_n$ in $Z$, the
\bfi{Gram matrix} of $z_1,\dots,z_n$, i.e., the $n\times n$-matrix $G$
with entries
\lbeq{e.gram}
G_{jk}=F(z_j,z_k),
\eeq
is Hermitian and positive semidefinite (resp. conditionally
semidefinite). In particular, every function of positive type is
conditionally positive.

The basic intuition for the above definition comes from the following
examples. (Note that $z$ and $z'$ are unrelated points.)

\begin{prop}\label{p.pos}
Let $Z$ be a subset of a Euclidean space $\Hz$. Then the functions
$F,F',F'':Z \times Z\to\Cz$ defined by
\[
F(z,z'):=z^* z',~~~ F'(z,z'):=z'^*z,~~~ F''(z,z'):=\re z^*z'
\]
are of positive type.
\end{prop}

\bepf
Let $G,G',G''$ be the Gram matrices computed with $F,F',F''$,
respectively. Clearly, $G$ is Hermitian; it is positive
semidefinite since
\[
u^*Gu=\D\sum_{j,k}\ol u_j z_j^*z_k u_k
=\Big\|\sum_kz_k u_k\Big\|^2\ge 0.
\]
$G'=\ol G$ and $G''=\half(G+\ol G)$ are easily seen to be
Hermitian and positive semidefinite, too.
\epf

The Moore--Aronszejn theorem (Theorem \ref{t.Moore} below) provides a
converse of Proposition \ref{p.pos}.

\begin{prop}\label{p.condP0}
If $F:Z\times Z\to \Cz$ is conditionally positive then, for any function
$f:Z\to\Cz$ and any $\gamma\ge 0$, the function $\wt F:Z\times Z\to\Cz$
defined by
\lbeq{e.Fgamma}
\wt F(z,z'):=\ol{f(z)}+f(z')+\gamma F(z,z') \for z,z'\in Z
\eeq
is conditionally positive.
\end{prop}

\bepf
Let $G,\wt G$ be the Gram matrices computed with $F$ and $\wt F$,
respectively. Clearly, $\wt G$ is Hermitian, and
\[
\wt G_{jk}=\ol{f(z_j)}+f(z_k)+\gamma G_{jk},
\]
hence $\D\sum_\ell u_\ell=0$ implies
\[
u^*\wt Gu
=\sum_{j,k}\ol u_j (\ol{f(z_j)}+f(z_k)+\gamma G_{jk}) u_k
=\gamma \D\sum_{j,k}\ol u_j G_{jk}u_k
=\gamma u^*Gu\ge 0.
\]
Thus $\wt G$ is conditionally semidefinite.
\epf

\begin{prop}\label{p.condP}
Let $Z$ be a subset of a Euclidean space $\Hz$. Then for any function
$g:Z\to\Cz$, the function $\wt F:Z\times Z\to\Cz$ defined by
\lbeq{e.Fnorm}
\wt F(z,z'):=\ol{g(z)}+g(z')-\|\ol z-z'\|^2 \for z,z'\in Z
\eeq
is conditionally positive.
\end{prop}

\bepf
This follows from Propositions \ref{p.pos} and  \ref{p.condP0} since
$\wt F(z,z')=\ol{f(z)}+f(z')+F''(z,z')$, where $f(z)=g(z)-\|z\|^2$.
\epf

For appropriate converses of Propositions \ref{p.condP0} and
\ref{p.condP} see the theorems bySchoenberg ansd Menger in
Section \ref{ss.SM} below.

\subsection{Constructing functions of positive type}
\label{ss.Constructions}

In this subsection we discuss a toolkit for the construction of such
explicit functions of positive type from simpler ingredients.
We provide a number of constructions that allow one to verify
positivity properties. For further constructions and numerous examples
in the form of exercises see \sca{Berg} et al. \cite{BerCR}.

\begin{prop}\label{p.basic}
For every family $\phi_z$ ($z\in Z$) of vectors $\phi_z$ in a Euclidean
vector space $\Hz$, the function $F$ defined by
\[
F(z,z'):=\<\phi_z,\phi_{z'}\>
\]
is of positive type.
\end{prop}

\bepf
The corresponding matrix $G$ from \gzit{e.gram} is clearly
Hermitian, and
\[
x^*Gx=\sum_{j,k} \ol x_jG_{jk}x_k
=\sum_{j,k}\ol x_j\<\phi_{z_j},\phi_{z_k}\>x_k
=\Big|\sum_k x_k\phi_{z_k}\Big|^2\ge 0.
\]
\epf

Basic examples of functions of positive type arise from the above
constructions by choosing the family of $\phi_z$ in such a way that
their inner products can be expressed in closed form.
Others come from a number of constructions which modify or combine
functions of positive type.

\begin{prop}\label{p.cons}~\\
(i) Every positive semidefinite Hermitian form on a complex vector
space $Z$ is of positive type.

(ii) If $F$ is of positive type over $Z$ and $Y\subseteq Z$ then
the \bfi{restriction} $F|_{Y}$ of $F$ to $Y\times Y$
is of positive type.

(iii) If $F_0$ is of positive type over $Z_0$ and $u:Z\to Z_0$ then
\[
F(z,z'):=F_0(u(z),u(z'))
\]
is of positive type.

(iv) If $F$ is of positive type over $Z$, $\gamma>0$, and
$\nu:Z\to\Cz$ then
\[
F'(z,z'):=\gamma \ol{\nu(z)}F(z,z')\nu(z')
\]
is of positive type. In particular, if $F(z,z)> 0$ for all $z$ then
the \bfi{normalization} $F_\norm$ of $F$, defined by
\[
F_\norm(z,z'):=\frac{F(z,z')}{\sqrt{F(z,z)F(z',z')}},
\]
is of positive type, and satisfies $F_\norm(z,z)=1$ for all $z$.

(v) If $L$ is a countable set and each $F_\ell$ ($\ell\in L$) is of
positive type over $Z$ then, for arbitrary positive weights $w_\ell$
for which
\[
F(z,z'):=\sum_{\ell\in L} w_\ell F_\ell(z,z')
\]
is everywhere defined, $F$ is of positive type.

(vi) Let $Z$ be the disjoint union of a family of sets $Z_\ell$ indexed
by $\ell\in L$. If $F_\ell:Z_\ell\times Z_\ell\to \Cz$ is of positive
type for all $\ell\in L$ then the function $F:Z\times Z\to \Cz$ defined
by
\[
F(z,z'):=\cases{F_\ell(z,z') & if $z,z'\in Z_\ell$,\cr
                0         & otherwise,}
\]
is of positive type.

(vii) If the $F_\ell$ ($\ell=0,1,2\ldots$) are of positive type over
$Z$ and the limit
\[
F(z,z'):=\D\lim_{\ell\to\infty} F_\ell(z,z')
\]
exists for  $z,z'\in Z$ then $F$ is of positive type.

(viii) If $\mu$ is a positive measure on a set $L$ and each $F_\ell$
($\ell\in L$) is of positive type over $Z$ then
\[
F(z,z'):=\int_L d\mu(\ell) F_\ell(z,z'),
\]
if everywhere defined, is of positive type.

(ix) Let $F_0:Z_0\times Z_0\to\Cz$ be of positive type, let $d\mu$ be
a positive measure on a set $L$. If $u:Z\times L\to Z_0$ is such
that the integral
\[
F(z,z')(\ell):=\int_L d\mu(\ell')F_0(u(z,\ell),u(z',\ell'))
\]
exists for all $\ell\in L$ and $z,z'\in Z$ then $F$ is of positive type
on $Z$.
\end{prop}

\bepf
(i)--(viii) are straightforward, and (ix) follows from (iii) and (viii).
\epf

Note that many examples of interest are analytic in the second
argument. Unfortunately, this property does not persist under
normalization as in Proposition \ref{p.cons}(iv).

It is easily checked that all constructions of Proposition \ref{p.cons}
produce conditionally positive functions when the ingredients are
only required to be conditionally positive rather than of positive type.

\begin{thm}\label{p.cons2} (\sca{Schur} \cite{Schur})~\\
(i) If $F_1$ is of positive type on $Z_1$ and  $F_2$ is of positive
type on $Z_2$ then
\[
F((z_1,z_2),(z_1',z_2')):=F_1(z_1,z_1')F_2(z_2,z_2')
\]
is of positive type on $Z=Z_1\times Z_2$.

(ii) If $F_1$ and $F_2$ are of positive type then the pointwise product
\[
F(z,z'):=F_1(z,z')F_2(z,z')
\]
is of positive type.
\end{thm}

\bepf
(i) For $t=1,2$, the Gram matrix $G_t$ of $z_{t1},\ldots,z_{tn}$
computed with respect to $F_t$ is positive semidefinite, hence has a
Cholesky factorization $G_t=R_t^*R_t$.
The Gram matrix of $(z_{11},z_{21}),\ldots,(z_{11},z_{21})$ computed
with respect to $F$ has entries
\[
\bary{lll}
G_{jk}&=&G_{1jk}G_{2jk}
=\Big(\D\sum_\ell \ol{R_{1\ell j}}R_{1\ell k}\Big)
 \Big(\sum_m \ol{R_{2m j}}R_{2m k}\Big)\\
&=&\D\sum_{\ell,m}\ol{R_{1\ell j}R_{2m j}}R_{1\ell k}R_{2m k},
\eary
\]
so that
\[
u^*Gu=\sum_{j,k}\ol u_j G_{jk} u_k =
\D\sum_{\ell,m}\Big|\sum_j u_jR_{1\ell j}R_{2m j}\Big|^2\ge 0.
\]
Thus $G$ is positive definite, proving that $F$ is of positive type.

(ii) follows from (i) and Proposition \ref{p.cons}(iii) by mapping to
the diagonal.
\epf

\begin{thm}\label{t.powers}~\\
(i) All pointwise powers
\[
F^n(z,z'):=F(z,z')^n~~~(n=1,2,\dots)
\]
of a function $F$ of positive type are of positive type.

(ii) If $F$ is of positive type then for any $\beta\ge 0$, the function
$F_\beta$ defined by
\[
F_\beta(z,z'):=e^{\beta F(z,z')}
\]
is of positive type, too.

(iii) Write $B(0;1):=\{x\in \Cz \mid |x|<1\}$ for the open complex
unit disk. If $F$ is of positive type and $|F(z,z')|<c<\infty$ for all
$z,z'\in Z$ then
\[
F_\inv(z,z'):=\frac{1}{c-F(z,z')}
\]
is of positive type, too. (This is related to Nevanlinna--Pick
interpolation theory; cf. \sca{Agler \& McCarthy} \cite{AglM}.)
\end{thm}

\bepf
(i) follows from Theorem \ref{p.cons2}(ii) by induction.
(ii) and (iii) then follow from Proposition \ref{p.cons}(v) since
$e^x=\D\sum_0^\infty \frac{x^n}{n!}$ for $x\in \Cz$ and
$\D\frac{1}{c-x}=\sum_0^\infty\frac{x^n}{ c^{n+1}}$ for $|x|<c$,
and constant functions with positive values are of positive type.
\epf

This theorem is related to the Berezin--Wallach set discussed in
Section \ref{ss.BWset}.

\section{Coherent spaces and their quantum spaces} \label{s.cohSp}

\subsection{Coherent spaces} \label{ss.cohSp}

Let $Z$ be a nonempty set. A \bfi{coherent product} on $Z$ is a
function $K:Z\times Z\to\Cz$ of positive type.\footnote{
One obtains the more general concepts of \bfi{semicoherent products}
and \bfi{semicoherent spaces}
by weakening the requirement of having positive type to the requirement
that the supremum $\fns{ns}(Z)$ of the number of negative eigenvalues
of Gram matrices constructed from $K$ is finite.
Much of the subsequent theory remains valid, but the inner products
need no longer be positive semidefinite and the quantum spaces
discussed below become Pontryagin spaces with $\fns{ns}(Z)$ negative
squares; cf. \sca{Alpay} et al. \cite{AlpDR}.
In the present paper, this generalization is not considered further.
} 
A \bfi{coherent space} is a nonempty set $Z$ with a distinguished
coherent product $K:Z\times Z\to\Cz$. We regard the same set with
different coherent products as different coherent spaces.

\begin{expls}\label{ex.coh}~\\
(i)
Any subset $Z$ of a Euclidean space is a coherent space with
coherent product
\[
K(z,z'):=z^*z'.
\]
(ii)
Any subset $Z'$ of a coherent space $Z$ is again a coherent
space, with the coherent product inherited from $Z$ by restriction.

(iii)
For practical applications, it is often important that the coherent
products are given as explicit expressions $K(z,z')$ with which one
can work analytically, or at least expressions which can be efficeintly
approximated numerically.
The easiest way to construct such expressions is by using one of the
many constructions from Subsection \ref{ss.Constructions}.
\end{expls}

Many interesting examples will appear in other papers on this
subject, starting with \sca{Neumaier \& Ghaani Farashahi}
\cite{NeuF.cohQuant}. We just give one particularly important example.
As we shall see in \cite{NeuF.cohQuant}, the corresponding quantum
spaces are the Fock spaces upon which quantum field theory is based.

\begin{expl}\label{ex.Klauder}
Let $V$ be a Euclidean space. In a notation where pairs are denoted by
square brackets, we write
\[
z:=[z_0,\z]\in \Cz\times V.
\]
for the elements of $Z=\Cz\times V$. Since
\lbeq{e.oscF}
F(z,z'):=\ol z_0 +z_0'+\z^*\z'
\eeq
is conditionally positive by Proposition \ref{p.condP0}. Hence
Theorem \ref{t.divis} below implies that
\lbeq{e.Kosc}
K(z,z'):=e^{\ol z_0 +z_0'+\z^*\z'}
\eeq
is a coherent product, with respect to which $Z$ is a
coherent space. We call this coherent space the \bfi{Klauder space}
over $V$ and denote it by $Kl[V]$.
(For $V=\Cz$, the associated coherent states were first discussed in
\sca{Klauder} \cite[p.1062]{Kla63}.)
We shall discuss Klauder spaces in more detail in
\sca{Neumaier \& Ghaani Farashahi} \cite{NeuF.cohQuant}.
\end{expl}

In particular, coherent spaces generalize Euclidean spaces, and the
coherent product $K(z,z')$ generalizes the Hermitian inner product
$z^*z'$, but in general no linear structure is assumed on $Z$.
This is similar to the way how metric spaces generalize the distance
in Euclidean spaces without keeping their linear structure.

We draw some simple but useful general consequences.
The Hermiticity of the Gram matrix of $z,z'$ gives
\lbeq{cp2}
\ol{K(z,z')}=K( z', z).
\eeq
Since the diagonal elements of a Hermitian positive semidefinite
matrix are real and nonnegative,
\lbeq{cp1}
K( z,z)\ge 0\Forall z\in Z.
\eeq
In particular, we may define the \bfi{length} of $z\in Z$ to be
\lbeq{e.nz}
n(z):=\sqrt{K( z,z)}\ge 0.
\eeq
Since every principal submatrix of a Hermitian positive semidefinite
matrix has real nonnegative determinants, the determinants of size 2
lead to
\lbeq{e.gram2}
|K(z,z')|^2 \le K(z, z)K( z',z').
\eeq
Taking square roots gives the \bfi{coherent Cauchy--Schwarz inequality}
\lbeq{e.cohCS}
|K(z,z')| \le n( z)n(z').
\eeq
This allows us to define the \bfi{angle} between two points $z,z'\in Z$
of positive length by
\lbeq{e.cohAngle}
\angle(z,z'):=\arccos \frac{|K(z,z')|}{n( z)n(z')} \in [0,\pi[.
\eeq

\subsection{Quantum spaces} \label{s.quant}

Let $Z$ be a coherent space.
A \bfi{quantum space} of $Z$ is a Euclidean space $\Qz(Z)$
spanned by (i.e., consisting of all finite linear combinations of) a
distinguished set of vectors $|z\>$ ($z\in Z$) satisfying
\[
\<z|z'\> :=\<z|\,|z'\>= K( z,z') \for z,z'\in Z,
\]
with the linear functionals\footnote{
With this convention, $\<z|$ is a linear functional mapping
$\psi\in\Qz^\*(Z)$ to $\<z|psi$, while $|z\>\in\Qz(Z)$ is an antilinear
functional mapping $\psi\in\Qz^\*(Z)$ to $\psi^*|z\>$.
} 
\[
\<z|:=|z\>^*
\]
acting on $\Qz^\*(Z)$.
Thus there is a distinguished map from $Z$ to $\Qz(Z)$ mapping $z$ to
the vectors $|z\>$ ($z\in Z$); these are called the
\bfi{coherent states} of $Z$ in $\Qz(Z)$. In this paper, we use this
Dirac bra/ket notation {\em only} for coherent states and their
adjoints.

We call the completion $\ol\Qz(Z):=\ol{\Qz(Z)}$ of a quantum space the
corresponding \bfi{completed quantum space} of $Z$. The corresponding
\bfi{augmented quantum space} is the antidual $\Qz^\*(Z):=\Qz(Z)^\*$.
We have
\[
\Qz(Z) \subseteq \ol\Qz(Z) \subseteq \Qz^\*(Z).
\]
If the quantum space is infinite-dimensional, $\Qz(Z)$ is usually a
proper subspace of the Hilbert space $\ol\Qz(Z)$.
By definition of the weak-* topology of $\Qz^\*(Z)$,
$\psi_\ell\in\Qz^\*(Z)$ converges to $\psi\in\Qz^\*(Z)$ iff
$\<z|\psi_\ell\to\<z|\psi$ for all $z\in Z$.

\begin{prop}\label{p.cohSp}~\\
(i) Let $\Hz$ be a Euclidean space. Then for any set $Z$ and any mapping
$c:Z\to\Hz$,
\lbeq{e.Kcc}
K(z,z'):=c(z)^*c(z')
\eeq
defines a coherent product on $Z$ that turns $Z$ into a coherent space
whose quantum space $\Qz(Z)$ is the space consisting of the
finite linear combinations of coherent states $|z\>:=c(z)$.
($\Qz(Z)$ is usually a proper subspace of $\Hz$.)

(ii) Conversely, every coherent product can be written in the form
\gzit{e.Kcc} such that the coherent states are given as $|z\> = c(z)$.
\end{prop}

\bepf
(i) follows by combining Example \ref{ex.coh}(i) with the definition
of the quantum space. To see (ii), take $\Hz=\Qz(Z)$ and
define $c(z):=|z\>$.
\epf

\begin{thm}\label{t.qSpace}
Every coherent space $Z$ has a quantum space $\Qz(Z)$.
It is unique up to an isomorphism that maps coherent states with the
same label to each other.
\end{thm}

\bepf
By definition of a coherent space, the coherent product $K$ is of
positive type. Hence the Moore--Aronszejn theorem (Theorem
\ref{t.Moore} below) applies and provides
a Hilbert space $\ol\Qz$. If we define the coherent states $|z\>:=q_z$
and their adjoints $\<z|:=q_z^*$, we find from \gzit{cp5} below that
\[
\<z|z'\>=\<q_z,q_{z'}\>=K(z,z').
\]
Thus the space $\Qz$ consisting of the finite linear combinations of
coherent states is a quantum space.
If $\Qz$ and $\Qz'$ are quantum spaces for $Z$ with coherent
states $|z\>$ and $|z\>'$, respectively, then
\[
I(\phi):=\sum a_k|z_k\>' \iif \phi=\sum a_k|z_k\>
\]
defines a map $I:\Qz\to\Qz'$. Indeed, if $\phi=\sum b_k|z_k\>$ is
another representation of $\phi$ then
\[
\sum a_k K( z,z_k)=\sum a_k\<z|z_k\>=\<z|\phi=\sum b_k\<z|z_k\>
=\sum b_k K( z,z_k).
\]
Thus $\phi':=\sum b_k|z_k\>'$ satisfies
\[
{}'\<z|\phi'=\sum b_k{}'\<z|z_k\>'=\sum b_k K( z,z_k)
=\sum a_k K( z,z_k)=\sum a_k{}'\<z|z_k\>'={}'\<z|I(\phi)
\]
for all $z\in Z$, whence $\phi'=I(\phi)$. This map is easily seen to be
an isomorphism.
\epf

Note that any linear map from a quantum space of a coherent space
into $\Cz$ is continuous, and any linear map from a quantum space of
a coherent space into its antidual is continuous, too.

\bigskip

Let $Z,Z'$ be coherent spaces.
A \bfi{morphism} from $Z$ to $Z'$ is a map $\rho:Z\to Z'$ such that
\lbeq{mor.coh}
K'(\rho(z),\rho(z))=K(z,w) \for  z,w\in Z;
\eeq
if $Z'=Z$, $\rho$ is called an \bfi{endomorphism}.
Two coherent spaces $Z$ and $Z'$ with coherent products $K$ and $K'$,
respectively, are called \bfi{isomorphic} if there is
a bijective morphism $\rho:Z\to Z'$.  In this case we write $Z\cong Z'$
and we call the map $\rho:Z\to Z'$ an \bfi{isomorphism} of the coherent
spaces. Clearly, $\rho^{-1}:Z'\to Z$ is then also an isomorphism.
If $Z'=Z$ and $K'=K$ we call $\rho$ an \bfi{automorphism} of $Z$.
Automorphisms are closely related to the more general concept of
coherent maps, introduced in
\sca{Neumaier \& Ghaani Farashahi} \cite{NeuF.cohQuant}.

\begin{prop}\label{main.iso.p}
Let $Z,Z'$ be coherent spaces and $\rho:Z\to Z'$ be an isomorphism.
Then,

(i) $K(\rho^{-1}(z'),\rho^{-1}(w'))=K'(z',w')$ for all $z',w'\in Z'$.

(ii) $K'(z',\rho(z))=K(\rho^{-1}(z'),z)$ for all $z\in Z$ and
$z'\in Z'$.
\end{prop}
\bepf
(i) is straightforward.

(ii) Let $z\in Z$ and $z'\in Z'$. Then
$K'(z',\rho(z))=K'(\rho(\rho^{-1}(z')),\rho (z))=K(\rho^{-1}(z'),z)$.
\epf

\begin{prop}
Let $Z$ be a coherent space with coherent product $K$ and $Z'$ be
an arbitrary set. Then for any map $\rho :Z'\to Z$,
\[
K'(z,z'):=K(\rho z,\rho z') \for  z,z'\in Z'
\]
defines a coherent product on $Z'$. This turns $Z'$ into a coherent
space with respect to which $\rho $ is a morphism.
\end{prop}

\begin{prop}
Let $Z,Z'$ be isomorphic coherent spaces. Then any two quantum spaces
$\Qz(Z)$ of $Z$ and $\Qz(Z')$ of $Z'$ are isometric Euclidean spaces.
\end{prop}

\bepf
Let $Z,Z'$ be isomorphic coherent spaces. Let $\Qz(Z)$ and $\Qz(Z')$ be
quantum spaces of $Z$ and $Z'$, respectively.
Let $\rho:Z\to Z'$ be an isomorphism of coherent spaces. We define
the map $T_\rho:\Qz(Z)\to\Qz(Z')$ given by
\[
T_\rho\Big(\sum_k c_k|z_k\>\Big):=\sum_kc_k|\rho (z_k)\>'
\Forall \sum_kc_k|z_k\>\in\Qz(Z).
\]
Now
\[
\bary{lll}
\D\Big\|T_\rho\Big(\sum_k c_k|z_k\>\Big)\Big\|_{\Qz(Z')}^2
&=&\D\Big\|\sum_kc_k|\rho (z_k)\>'\Big\|^2_{\Qz(Z')}=
\sum_k\sum_j\ol{c_k}c_jK'(\rho (z_k),\rho (z_j))\\
&=&\D\sum_k\sum_j\ol{c_k}c_jK(z_k,z_j)
=\D\Big\|\sum_k c_k|z_k\>\Big\|_{\Qz(Z)}^2.
\eary
\]
This implies that $T_\rho$ is a well-defined isometry. Since $\rho$ is
surjective, $T_\rho$ is surjective as well. Thus, $T_\rho$ is an
isomorphism.
\epf

\subsection{Constructing vectors in the augmented quantum space}
\label{ss.states}

By definition, all vectors in $\Qz^\*(Z)$ can be constructed as weak-*
limits of nets in $\Qz(Z)$, and hence by the following construction.

\begin{prop}
A net $\psi_\ell$ in $\Qz(Z)$ is weak-* convergent iff, for all
$z\in Z$, the inner product $\<z|\psi_\ell$ converges. In this case,
the limit $\psi=\lim\psi_\ell \in\Qz^\*(Z)$ is characterized by
\lbeq{e.lim}
\psi(|z\>):=\<z|\psi=\lim\<z|\psi_\ell \for z\in Z.
\eeq
\end{prop}

\bepf
By definition, weak-* convergence to $\psi$ says that
$\psi_\ell(\phi)\to\psi(\phi)$ for all $\phi\in\Qz(Z)$. In particular,
$\<z|\psi_\ell=\psi_\ell(|z\>)$ converges to $\psi(|z\>)$, and
\gzit{e.lim} holds. Conversely, suppose that the limit \gzit{e.lim}
exists for all $z\in Z$. Any $\phi\in \Qz(Z)$ can be written as a
finite linear combination $\phi=\D\sum_k\alpha_k|z_k\>$, hence
$\psi_\ell(\phi)=\D\sum_k\alpha_k\psi_\ell(|z_k\>)
=\sum_k\<z_k|\psi_\ell$ converges. Thus the net $\psi_\ell$ is weak-*
convergent.
\epf

Let $X$ be an open subset of a finite-dimensional complex vector space.
We call a map $u:X\to Z$ \bfi{smooth} if, for each $z\in Z$,
$K(z,u(x))$ is $C^\infty$ as a function of $x\in X$, and
\bfi{strongly smooth} if, in addition, $K(u(x),u(y))$ is $C^\infty$ in
$(x,y)\in X\times X$. (With this notions of smoothness, $Z$ becomes in
two ways a diffeological space; cf. \sca{Iglesias-Zemmour} \cite{IglZ}).

\begin{thm}
Let $u:X\to Z$ be a smooth map and let $A$ be a linear differential
operator on $C^\infty(X)$.  Then:

(i) For any $x\in X$, there is a unique state $\psi_{u,A,x}\in\Qz^\*(Z)$
such that
\lbeq{e.zpsi}
\<z|\psi_{u,A,x}=A(x)K(z,u(x)).
\eeq
(ii) For fixed $u,x$, the map $A\to\psi_{u,A,x}$ is linear.

(iii) If $u$ is strongly smooth then $\psi_{u,A,x}\in\ol\Qz(Z)$.

(iv) If $u:X\to Z$ and $v:Y\to Z$ are strongly smooth, if $A,B$ are
linear differential operators on $C^\infty(X)$ and $C^\infty(Y)$,
respectively, and $x\in X,y\in Y$ then
\[
\psi_{u,A,x}^*\psi_{v,B,y}=\ol{A(x)\ol{B(y)K(u(x),u(y))}}.
\]
\end{thm}

\bepf
(i) Every linear differential operator $A$ on $C^\infty(X)$ can be
written as a limit of a sequence of finite linear combinations of
function values,
\[
A(x)f(x)=\lim_{\ell\to\infty}\sum_k\alpha_{\ell k}f(x+h_{\ell k}),
\]
obtainable by replacing each derivative by a limit of difference
quotients. We apply this to the function $f\in C^\infty(X)$ defined by
$f(x):=K(z,u(x))$ and find that the sequence of vectors
$\psi_\ell:=\D\sum_k\alpha_{\ell k}|u(x+h_{\ell k})\>$ satisfies
\[
\lim_\ell\<z|\psi_\ell
=\lim_\ell\sum_k\alpha_{\ell k}\<z|u(x+h_{\ell k})\>
=\lim_\ell\sum_k\alpha_{\ell k}K(z,u(x+h_{\ell k}))
=\lim_\ell A(x)K(z,u(x)).
\]
Thus $\psi:=\lim\psi_\ell$ exists and satisfies \gzit{e.zpsi}.

(ii) is straightforward.

(iii) and (iv) are proved similarly. \at{details?}
\epf

The above expressions for inner products make many calculations easy
that are difficult in Hilbert spaces whose inner product is defined
through a measure. In particular, the quantum spaces of coherent spaces
with an easily computable coherent product can be studied efficiently
without measures or integrations, in terms of the explicit coherent
product and differentiation only.

\subsection{Some examples}\label{ss.ex}

We now give a long list of basic examples of coherent spaces exhibiting
the flavor of the relations to other fields of mathematics and science.
As indicated in the introduction, this is just the tip of an iceberg;
many other coherent spaces will be discussed in subsequent papers of
this series.

The first group of examples arises in applications to quantum mechanics.
For the physical background see, e.g.,
\sca{Neumaier \& Westra} \cite{NeuW}.

\begin{expls}\label{ex.simpleEx}

The simplest instances of  coherent spaces are the spaces
formed by the subsets $Z$ of $\Cz^n$ which are closed under conjugation
and scalar multiplication, with one of the  coherent products
\lbeq{e.cohps1}
K(z,z'):=\cases{1 & if $z'=\ol z$,\cr
                0 & otherwise,}
\eeq
\lbeq{e.cohps2}
K(z,z'):=z^*z',
\eeq
\lbeq{e.cohps3}
K(z,z'):=(z^*z')^{2j} ~~~~~~(\mbox{$j=0,\half,1,\frac{3}{2},\dots$}),
\eeq
\lbeq{e.cohps4}
K(z,z'):=e^{(z^*z'-\half\|z\|^2-\half\|z'\|^2)/\hbar},
\eeq
where $\hbar$ is a positive real number. In the applications to quantum
mechanics, $\hbar$ is the \bfi{Planck constant}.
The axioms are easily verified using the constructions of Proposition
\ref{p.cons} and Theorem \ref{t.powers}.

(i)
$Z=\Cz^n$ with the  coherent product \gzit{e.cohps1}
corresponds to the phase space of a classical system of $n$ oscillators,
with $n$ position and $n$ momentum degrees of freedom, via the
identification
\[
z=q+i p,~~~ q=\re z,~~~ p=\im z.
\]
In the corresponding quantum space, the associated coherent states are
orthonormal basis vectors, indexed by the phase space points.

(ii)
The unit sphere $Z$ in $\Cz^2$ with the  coherent product
\gzit{e.cohps1} corresponds to the phase space of a
\bfi{classical spin}, such as a polarized light beam or a spinning top
fixed at its point.

(iii)
The unit ball $Z$ in $\Cz^2$ with the  coherent product
\gzit{e.cohps1} corresponds to the classical phase space of
(monochromatic) \bfi{partially polarized light}.

(iv)
$Z=\Cz^n$ with the  coherent product \gzit{e.cohps2} has as
quantum space the Hilbert space $\Cz^n$ of an $n$-level quantum system.
The associated coherent states are all state vectors.

(v)
The unit sphere $Z$ in $\Cz^2$ with the  coherent product
\gzit{e.cohps3} corresponds to the Poincar\'e sphere (or Bloch sphere)
representing a single quantum mode of an atom with spin $j$, or for
$j=1$ the polarization of a single photon mode.
The corresponding quantum space has dimension $2j+1$. The associated
coherent states are the so-called \bfi{spin coherent states}.
(This example shows that a given set $Z$ may carry more than one
interesting coherent product, resulting in different coherent spaces
with nonisomorphic quantum spaces.)
For $j\to\infty$, the space degenerates into the coherent space of
a classical spin.

(vi)
$Z=\Cz^n$ with the  coherent product \gzit{e.cohps4} has as
quantum space the bosonic Fock space with $n$ degrees of freedom,
corresponding to $n$ independent harmonic oscillators. The associated
coherent states are the so-called \bfi{Glauber coherent states}.
In the so-called classical limit $\hbar\to 0$ (which can be taken
mathematically, though not in reality), the space degenerates into the
coherent space of a classical system with $n$ spatial degrees of
freedom.
\end{expls}

We note that for \gzit{e.cohps4}, the power construction from Theorem
\ref{t.powers}(i) just amounts to a replacement of $\hbar$ by $\hbar/n$.
Therefore the classical limit amounts here to applying the power
construction for arbitrary $n$ and considering the limit $n\to\infty$.
Generalizing this to arbitrary coherent spaces provides a general
definition of the \bfi{classical limit}, even when $\hbar$ does not
appear in the coherent product. For example, the power construction
applied to \gzit{e.cohps2} produces \gzit{e.cohps3} with $2j=n$; thus
the classical limit amounts here to the limit of infinite spin. The
classical limit and related semiclassical expansions are investigated
in general in a later paper of this series (\sca{Neumaier}
\cite{Neu.cohSpec}).

We also note that in order that a coherent product results, $\hbar$
can take in \gzit{e.cohps4} any positive value, while in
\gzit{e.cohps3}, $2j$ must be a nonnegative integer. (The latter is
already needed in order that the power is unambiguously defined.)
This phenomenon is captured through the concept of a Berezin--Wallach
set (see Subsection \ref{ss.BWset} below).

\begin{expl}\label{ex.Kmin}
The set $Z=\Rz_+$ of positive real numbers is a real coherent
space with trivial conjugation for any of the coherent
products
\[
K(z,z')=\min(z,z'),
\]
\[
K(z,z')=(z+z')^{-1}.
\]
(i) In the first case, a completed quantum space is $L^2(\Rz_+)$ with
coherent states
\[
k_z(z')=\cases{1 & if $z'\le z$,\cr
               0 & otherwise.}
\]
(ii) In the second case, a completed quantum space is $L^2(\Rz_+)$ with
coherent states
\[
k_z(z')=e^{-zz'}
\]
since
\[
\<k_z, k_{z'}\> = \int^\infty_{0+}\!\!dy\; \ol{k_z(y)} \, k_{z'}(y)
  = \int^\infty_{0}\!\!dy\; e^{-zy} e^{-z'y}
  = \int^\infty_{0}\!\!dy\; e^{-(z+z')y}
  = \frac{1}{z+z'}.
\]
\end{expl}

The following spaces are important not only in complex analysis but are
also relevant in quantum physics, for the analysis of quantum mechanical
scattering problems (\sca{de Branges \& Rovnyak}
\cite[Theorem 4]{deBraR}). Example \ref{ex.deB}(ii) below is relevant
in signal processing.

For a function $f:Z\subset\Cz\to\Cz$ we define its conjugate
$\ol f:Z\to\Cz$ by
\lbeq{e.cconj}
\ol f(z):=\ol{f(\ol z)}.
\eeq

\begin{expls}\label{ex.deB}~\\
(i) A \bfi{de Branges function} is an entire analytic function
$E:\Cz\to\Cz$ satisfying
\lbeq{e,dB}
|E(\ol z)|<|E(z)| \iif \im z>0.
\eeq
With the  coherent product
\[
K(z,z'):=\cases{
  \ol E'(\ol z)E(z')-E'(\ol z)\ol E(z')   & if $z'=\ol z$,\cr
                                 ~~~      &~\cr
\D\frac{\ol E(\ol z)E(z')-E(\ol z)\ol E(z')}{2i(\ol z-z')} & otherwise,}
\]
$Z=\Cz$ is a  coherent space. A corresponding quantum space is the
subspace of $L^2(\Rz)$ spanned by the coherent states $q_z$, denoted by
$\mathcal{H}(E)$, defined by
\[
q_z(t)=\frac{K(\ol z,t)}{E(t)}
:=\lim_{\eps\downto 0} \frac{K(\ol z,t+i\eps)}{E(t+i\eps)}
\for t\in\Rz.
\]
(The denominator on the right is nonzero by \gzit{e,dB}.
The limit exists and is continuous as a function of $t$ since at an
$n$-fold zero $t$ of $E$, the function $K(\ol z,\cdot)$ has $t$ as
a zero of multiplicity at least $r$.)
Indeed, the formula $q_z^*q_{z'}=K(z,z')$ follows by evaluating the
integral expression for $q_z^*q_{z'}$ using the residue theorem.
For details see \sca{de Branges} \cite[Theorem 19, p.50]{deBra},
where the quantum space is more fully characterized.

(ii) $Z=\Cz$ is a coherent space with the coherent product
\[
K(z,z'):=\sinc(\ol z-z'),~~~
\sinc(z):=\cases{1         & if $z=0$,\cr
                 \sin(z)/z & otherwise.}
\]
This is the special case $E(z)=e^{-iz}$ of (i).

(iii) A \bfi{Schur function} (\sca{Schur} \cite{Schur18}) is an
analytic function $s$ from the open unit disk $B(0;1)$ in $\Cz$ into
its closure. With the  coherent product
\[
K(z,z'):=\frac{1-\ol{s(z)}s(z')}{1-\ol zz'}.
\]
$Z:=B(0;1)$ is a coherent space. Note that the inverse is defined since
$|\ol zz'|<1$. Coherence follows from results by
\sca{de Branges \& Rovnyak} \cite{deBraR.sum}. The corresponding
quantum spaces are the sub-Hardy spaces discussed by \sca{Sarason}
\cite{Sar}, also called \bfi{de Branges--Rovnyak spaces}; see the
recent survey by \sca{Ball \& Bolotnikov} \cite{BalB}.

(iv) The \bfi{Szeg\"o space} is the coherent space defined on the
open unit disk in $\Cz$,
\[
D(0,1):=\{z\in\Cz\mid |z|<1\},
\]
by the coherent product
\[
K(z,z'):=(1-\ol zz')^{-1},
\]
cf. \sca{Szeg\"o} \cite{Sze}. This example from 1911 is probably the
earliest nontrival explicit, nontrivial coherent product in the
literature. It is the special case $s=0$ of (iii); its quantum space is
the Hardy space on the unit disk. Coherence also follows directly
from Theorem \ref{t.powers}(iii).
\end{expls}

In general, unlike in these (and other simple) examples, there need not
be a simple realization of a quantum space in terms of an $L^2$ space
with respect to a suitable measure. Fortunately, such a description is
usually not needed in applications to physics since one can work
comfortably in the quantum space using only its defining properties.
This is one of the strengths of the concept of coherent spaces, as it
allows one to avoid the often cumbersome evaluation of integrals in
the computation of inner products.

\section{Nondegenerate and projective coherent spaces}

In this section we consider some desirable conditions a coherent space
may or may not have. Some of these conditions are satisfied in many
coherent spaces relevant for the applications.

\subsection{The distance}

Theorem \ref{t.qSpace} implies that, in a sense, coherent spaces are
just the subsets of Euclidean spaces.
However, separating the structure of a coherent space $Z$ from the
notion of a Euclidean space allows many geometric features to
be expressed in terms of $Z$ and the coherent product alone, without
direct references to the quantum space. The latter only serves as a
convenient tool for proving assertions of interest. For example, the
study of symmetry in \sca{Neumaier \& Ghaani Farashahi}
\cite{NeuF.cohQuant} benefits from this separation. Another example is
the distance function induced on $Z$ by the Euclidean distance, as in
the proof of Proposition \ref{p.d(z,z')} below. It will play an
important role in the study of coherent manifolds in
\sca{Neumaier \& Ghaani Farashahi} \cite{NeuF.cohMan}.

\begin{prop}
\label{p.d(z,z')}
(\sca{Parthasarathy \& Schmidt} \cite[Corollary 1.3/4]{ParS})\\
The \bfi{distance}
\lbeq{e.dist}
d(z,z'):=\sqrt{K( z,z)+K( z',z')-2\re K( z,z')},
\eeq
of two points $z,z'\in Z$ is nonnegative and satisfies the triangle
inequality. With \gzit{e.nz} we have
\lbeq{e.ndn}
|n(z)-n(z')|\le d(z,z')\le n(z)+n(z'),
\eeq
\lbeq{e.lip}
|K(y,z)-K(y',z')|\le d(y,y')n(z')+n(y)d(z,z').
\eeq
\end{prop}

\bepf
The expression under the square root of \gzit{e.dist} is
\lbeq{d.to.norm}
\<z|z\>+\<z'|z'\>-\<z|z'\>-\<z'|z\>=\Big\||z\>-|z'\>\Big\|^2,
\eeq
whence $d(z,z')$ is just the Euclidean distance between $|z\>$ and
$|z'\>$. This implies nonnegativity and the triangle inequality.
$n(z)$ is the length of $|z\>$, and \gzit{e.ndn} follows.
The Cauchy--Schwarz inequality gives
\[
|K(y,z)-K(y,z')|=\Big|\<y|\Big(|z\>-|z'\>\Big)\Big|\le n(y)d(z,z'),
\]
hence
\[
\bary{lll}
|K(y,z)-K(y',z')|&=|K(y,z)-K(y,z')+K(y,z')-K(y',z')|\\[3mm]
&\le|K(y,z')-K(y',z')|+|K(y,z)-K(y,z')|\\[3mm]
&\le d(y,y')n(z')+d(z,z')n(y).
\eary
\]
This proves \gzit{e.lip}.
\epf

We call a coherent space \bfi{nondegenerate} if $K(z'',z')=K(z,z')$
for all $z'\in Z$ implies $z'' = z$. Clearly, this is the case iff
the mapping from $Z$ to $\Qz(Z)$ that maps each $z\in Z$ to the
corresponding coherent state $|z\>$ is injective.

\begin{prop}
The distance map $d$ is a metric on $Z$ iff $K$ is nondegenerate on $Z$.
\end{prop}

\bepf
\gzit{d.to.norm} implies that $d(z,z')=0$ iff $|z\>=|z'\>$. Hence
$d$ is a metric on $Z$ iff $K$ is nondegenerate on $Z$.
\epf

The distance map $d$ is a quasimetric, hence it induces in the
standard way a topology on $Z$ called the \bfi{metric topology} and
denoted by $\tau_m$. There is a second topology on nondegenerate
coherent spaces $Z$, the \bfi{coherent topology} denoted by $\tau_c$,
defined by calling a net $z_\ell$ \bfi{coherently convergent} to $z$
iff $K(z_\ell,z')\to K(z,z')$ for all $z'\in Z$.
It can be readily checked that the coherent topology $\tau_c$ is at
least as fine as the metric topology $\tau_m$, because if $z_n\to z$
in the metric topology then $z_n\to z$ in the coherent topology, too.

\begin{thm}
In any coherent space, the metric topology is the weakest (coarsest)
topology in which $K$ is continuous.
\end{thm}

\bepf
We equip $Z\times Z$ with the product topology induced by the metric
topology on $Z$. Let $(z_n,z_n')$ be a convergent sequence to
$(z,z')\in Z\times Z$. Then $z_n\to z$ and $z_n'\to z'$ in the metric
topology. Hence the sequence of $n(z_n)$ is bounded. Thus
\[
|K(z_n,z_n')-K(z,z')|\le d(z_n,z)n(z')+d(z_n',z')n(z_n),
\]
which implies that $\lim_n K(z_n,z_n')=K(z,z')$.

Now let $\tau$ be any topology on the coherent space $Z$ such that
$K:Z\times Z\to\Cz$ is continuous. To prove that $\tau$ is at least as
fine as $\tau_m$ we assume that $w_n\to w$ in $Z$ with respect to
$\tau$. Since $K$ is continuous with respect to $\tau$ and
$K(w,w)=n(w)^2$, we find
\[
\bary{lll}
\lim_nd(w_n,w)
&=&\lim_n\sqrt{K(w_n,w_n)+n(w)^2-2\re K(w_n,w)}\\
&=&\sqrt{K(w,w)+n(w)^2-2\re K(w,w)}\\
&=&\sqrt{n(w)^2+n(w)^2-2\re n(w)^2}=0,
\eary
\]
which implies that $w_n\to w$ in $Z$ with respect to the metric
topology as well. Thus $\tau$ is at least as fine as $\tau_m$.
This implies that the metric topology is the weakest (coarsest)
topology in which $K$ is continuous.
\epf

\at{Are the following statements true?
\\
(i) The coherent topology is the strongest (finest) topology in which
$K$ is continuous.
\\
(ii) The coherent topology is Hausdorff iff $K$ is nondegenerate
over $Z$.
\\
(iii) The metric topology is Hausdorff iff $K$ is nondegenerate
over $Z$.
} 

\subsection{Normal coherent spaces}

We call a coherent space \bfi{normal} if
\[
\cases{K(z,z')=1   & if $z'=z$,\cr
       |K(z,z')|<1 & otherwise.}
\]
In a normal coherent space, coherent states have norm $1$, hence the
distance simplifies to
\lbeq{e.distN}
d(z,z'):=c\sqrt{1-\re K( z,z')},~~~c=\sqrt{2}.
\eeq
This distance was studied by \sca{Arcozzi} et al. \cite{ArcRSW} with
$c=1$ rather than the above value. \gzit{e.distN} implies that a normal
coherent space $Z$ is nondegenerate.

\begin{prop} \label{p.ScalCohProds}
Let $Z$ be a coherent space with coherent product $K$. Then, for any
function $\gamma:Z\to \Cz$, the set $Z$ with \bfi{scaled coherent
product}
\[
K_\gamma(z,z'):=\ol{\gamma(z)}K(z,z')\gamma(z')
\]
is also a coherent space.
\end{prop}

\bepf
The Gram matrix $G'$ of the scaled coherent product has entries
\[
G_{jk}:= K_\gamma(z_j,z_k)
=\ol{\gamma(z_j)}K(z_j,z_k)\gamma(z_k)
\]
and is clearly Hermitian. For any vector $u$, we define the vector $v$
with components $v_k:=\gamma(z_k)u_k$ and find
\[
u^*G'u=\sum_{j,k}\ol u_j \ol{\gamma(z_j)}K(z_j,z_k)\gamma(z_k) u_k
=\sum_{j,k}\ol v_jK(z_j,z_k)v_k \ge 0.
\]
Thus $G'$ is positive semidefinite.
\epf

\begin{prop}
Let $Z$ be a coherent space. If the coherent product is not
identically zero then there is a normal, coherent space $Z'$ such
that there is an isomorphism $\alpha:\Qz (Z)\to\Qz (Z')$ with
\[
 \{\alpha|z\>\mid z\in Z\}
       \subseteq \{ \lambda|z'\> \mid \lambda\in\Cz,~z'\in Z'\}.
\]
Thus any image of a coherent state of $Z$ is a multiple of some
coherent state of $Z'$.
\end{prop}

\bepf
If $K(z,z)=0$, the coherent state $|z\>$ vanishes by the
Cauchy--Schwarz inequality \gzit{e.CauchySchwarz}.
Thus we can delete such
points from $Z$. By scaling using Proposition \ref{p.ScalCohProds}, we
may assume that $K(z,z)=1$ without changing the Hilbert space.
Now the proof of the Cauchy--Schwarz inequality \gzit{e.CauchySchwarz}
shows that if $|K(z,z')|=1$ then the coherent states $|z\>$ and
$|z'\>$ differ by a phase only; so we may delete one of them without
changing the Hilbert space. The new coherent space is normal.
\epf

\subsection{Projective coherent spaces}

We call a coherent space $Z$ \bfi{projective} if there is a
\bfi{scalar multiplication} that assigns to each $\lambda\in\Cz^\times$
and each $z\in Z$ a point $\lambda z \in Z$ such that
\lbeq{e.projective}
K(z,\lambda z') ~=~ \lambda^e K(z,z')\Forall z,z'\in Z,
\eeq
for some $e\in\Zz\setminus\{0\}$ called the \bfi{degree}.
Note that a coherent space cannot be both normal and projective.
Example \ref{ex.simpleEx}(v) is projective of degree $e=2j$,
Example \ref{ex.Kmin}(i) and (ii) are projective of degree $1$ and $-1$,
respectively.

There are important degenerate projective spaces where the scalar
multiplication is not associative because it is not canonically defined.
An example are the Klauder spaces from Example \ref{ex.Klauder}, which
are projective of degree 1 with the scalar multiplication
\[
\lambda[z_0,\z]:=[z_0+\log \lambda,\z],
\]
using an arbitrary but fixed branch of $\log$. The need to restrict
to a fixed branch causes the associative law to be not valid
universally. On the other hand, we have:

\begin{prop}\label{non.deg.proj}
Let $Z$ be a nondegenerate and projective space. Then the scalar
multiplication is associative:
\lbeq{e.projective1}
\lambda(\mu z)=(\lambda\mu)z
\for \lambda,\mu\in\Cz^\times,~z\in Z,
\eeq
\end{prop}

\bepf
Let $z\in Z$ and $\lambda,\mu\in\Cz^\times$. For all $z'\in Z$, we have
\[
K(\lambda(\mu z),z')=\ol{\lambda}^eK(\mu z,z')
=\ol{\lambda}^e\ol{\mu}^eK(z,z')
=(\ol{\lambda\mu})^eK(z,z')=K((\lambda\mu)z,z').
\]
Now nondegeneracy of $K$ implies \gzit{e.projective1}.
\epf

\begin{prop}\label{projective.property}
Let $Z$ be a projective coherent space of degree $e$.
Then
\lbeq{e.projective.alt}
K(\lambda z,z') ~=~ \ol{\lambda}^e K(z,z')\Forall z,z'\in Z,
\eeq
\lbeq{e.projective2}
|\lambda z\>=\lambda^e|z\> \for\lambda \in\Cz^\times,~z\in Z,
\eeq
\lbeq{e.projective3}
K(z,\lambda z')=K(\ol{\lambda}z,z') \for\lambda \in\Cz^\times,~z\in Z.
\eeq
\end{prop}

\bepf
\gzit{e.projective.alt} follows from the definition and \gzit{cp2}.
To prove \gzit{e.projective2}, let $z\in Z$ and $\lambda\in\Cz$.
Then, for all $z'\in Z$,
\[
\<z'|\lambda z\>=K(z',\lambda z)=\lambda^e K(z',z)=\lambda^e\<z'|z\>.
\]
Finally, using (\ref{e.projective}) and (\ref{e.projective.alt}), we get
\[
K(z,\lambda z')=\lambda^e K(z,z')=K(\ol\lambda z,z').
\]
\epf

Formula \gzit{e.projective3} suggests that it might be fruitful to
consider more general maps $A:Z\to Z$ satisfying
\[
K(z,\lambda z')=K(\ol{\lambda}z,z') \for\lambda \in\Cz^\times,~z\in Z.
\]
Such maps are called \bfi{coherent maps} and are studied in detail in
\sca{Neumaier \& Ghaani Farashahi} \cite{NeuF.cohQuant} and many later
papers of this series. Invertible coherent maps are of fundamental
importance as they describe the symmetry group of a coherent space.

Any coherent space can be extended to a projective coherent space
without changing the quantum space. The idea of a projective extension
can be traced back to \sca{Klauder} \cite{Kla.III}.

\begin{prop}\label{p.proj}~\\
Let $Z$ be a coherent space and $e$ be a nonzero integer.
Then the \bfi{projective extension} $PZ:=\Cz^\*\times Z$ of degree $e$
is a projective coherent space with coherent product
\lbeq{e.projextensionK}
K_{\rm pe}((\lambda,z),(\lambda',z'))
~:=~ \ol{\lambda}^e\, K(z,z') \, \lambda'^e
\eeq
and scalar multiplication $\lambda'(\lambda,z):=(\lambda'\lambda,z)$.
The corresponding quantum spaces $\Qz (Z)$ and $\Qz (PZ)$ are
canonically isomorphic.
\end{prop}

\bepf
It is straightforward to check that $PZ$ with respect to the projective
extension kernel $K_{\rm pe}$ is a projective coherent space of degree
$e$. The map $T:\Qz(PZ)\to\Qz(Z)$ given by
\[
\D T\sum_{\ell}c_{\ell}|(\lambda_\ell,z_\ell)\>
:=\sum_\ell c_\ell\lambda_\ell^e|z_\ell\>
\Forall \sum_{\ell}c_{\ell}|(\lambda_\ell,z_\ell)\>\in\Qz(PZ),
\]
is well-defined and linear. Also, we have
\[
\bary{lll}
\D\Big\|T\sum_{\ell}c_{\ell}|(\lambda_\ell,z_\ell)\>\Big\|^2_{\Qz(Z)}
&=&\D\Big\|\sum_\ell c_\ell\lambda_\ell^e|z_\ell\>\Big\|^2_{\Qz(Z)}
=\D\sum_j\sum_k \ol{c_j}\,\ol{\lambda_j^e}c_k\lambda_k^eK(z_j,z_k)\\
&=&\D\sum_j\sum_k \ol{c_j}c_kK_{\rm pe}((\lambda_j,z_j),(\lambda_k,z_k))
=\D\Big\|\sum_\ell c_\ell|(\lambda_\ell,z_\ell)\>\Big\|^2_{\Qz(PZ)},
\eary
\]
which implies that $T$ is an isometric linear operator. Thus, $T$ is
injective as well. Let $\psi=\D\sum_\ell c_\ell|z_\ell\>\in\Qz(Z)$
with $c_\ell\not=0$ for all $\ell$. Then
$\phi:=\D\sum_\ell |(c_\ell^{-e},z_\ell)\>\in\Qz(PZ)$ with
$T\phi=\psi$. Thus, $T$ is an isomorphism.
\epf

\begin{cor}
Let $Z$ be a coherent space and $e$ be a nonzero integer.
Then $Z$ is projective of degree $e$ iff $P_eZ\cong Z$.
\end{cor}

\bepf
Let $Z$ be a projective space of degree $e\in\mathbb{Z}$. We then define
$\rho:P_eZ\to Z$ by $\rho(\lambda,z):=\lambda z$, for all
$(\lambda,z)\in P_eZ$. It is easy to check that $\rho:P_eZ\to Z$ is an
isomorphism. Hence $P_eZ\cong Z$. Conversely, suppose that
$P_eZ\cong Z$ and let $\rho:P_eZ\to Z$ be an isomorphism of coherent
spaces. Then, with multiplication defined by
$\lambda z:=\rho\lambda\rho^{-1}z$, $Z$ is projective of degree $e$.
Indeed,using Proposition \ref{main.iso.p}(ii) for $z,z'\in Z$, we have
\[
K(z,\lambda z')=K(z,\rho\lambda\rho^{-1}z')
=K_e(\rho^{-1}z,\lambda\rho^{-1}z')
=\lambda^eK_e(\rho^{-1}z,\rho^{-1}z')=\lambda^e K(z,z').
\]
\epf

\subsection{Nondegenerate coherent spaces}

\begin{prop} \label{p.equiv}~\\
Let $Z$ be a coherent space. Define on $Z$ an equivalence
relation $\equiv$ by
\[
z\equiv z' \iff K(z,z'')=K(z',z'') \Forall z''\in Z.
\]
Then the set $[Z]$ of equivalence classes
\[
[z]:=\{z'\in Z| z'\equiv z\}~~~(z\in Z)
\]
is a nondegenerate coherent space with the coherent product
\lbeq{e.equivK}
K([z],[z']) ~:=K(z,z')\ \Forall z,z'\in Z.
\eeq
The corresponding quantum spaces $\Qz (Z)$ and $\Qz ([Z])$ are
canonically isomorphic. In particular, if $Z$ is projective then $[Z]$
is projective, with scalar multiplication $\lambda[z]:=[\lambda z]$.
\end{prop}

\bepf
Let $Z$ be a coherent space and $z,z',w,w'\in Z$ with $[z]=[w]$ and
$[z']=[w']$. Then
\[
K([z],[z'])=K(z,z')=K(w,z')=K(w,w')=K([w],[w']).
\]
Thus, $K:[Z]\times [Z]\to\Cz$ is well-defined. It is straightforward to
check that $([Z],K)$ is a coherent space. Now let $z,w\in Z$ with
$K([z],[z'])=K([w],[z'])$ for all $z'\in Z$. Hence
\[
K(z,z')=K([z],[z'])=K([w],[z'])=K(w,z'),
\]
for all $z'\in Z$, giving $[z]=[w]$. Thus $[Z]$ is nondegenerate.
Let $T:\Qz(Z)\to\Qz([Z])$ be given by $\psi\to T\psi$, where
$T\psi:=\sum c_\ell|[z_\ell]\>$ for
$\psi=\sum c_\ell|z_\ell\>\in\Qz(Z)$. If $\psi=\sum c_\ell|z_\ell\>=0$
then, for all $w\in Z$,
\[
\<[w]|T\psi=\sum c_\ell\<[w]|[z_\ell]\>=\sum c_\ell K([w],[z_\ell])
=\sum c_\ell K(w,z_\ell)=0.
\]

Thus $T\psi=0$. Hence $T:\Qz(Z)\to\Qz([Z])$ is a well-defined linear
operator. Also, for $\psi\in\Qz(Z)$, we have
\[
\bary{lll}
\|T\psi\|^2=\sum_j\sum _k\ol{c_j}c_k K([z_j],[z_k])
=\sum_j\sum _k\ol{c_j}c_k K(z_j,z_k)=\|\psi\|^2,
\eary
\]
which implies that $T$ is an isometry, hence injective. It is
straightforward to see that $T$ is surjective as well. Hence $T$ is an
isomrphism.

If $Z$ is projective then $[Z]$ is projective with the same degree,
with scalar multiplication $\lambda[z]:=[\lambda z]$. Indeed, if $Z$ is
projective of degree $e$, we have
\[
K([z],\lambda[z'])=K([z],[\lambda z'])=K(z,\lambda z')=\lambda^eK(z,z')
=\lambda^e K([z],[z']),
\]
for all $z,z'\in Z$ and $\lambda\in\Cz^\*$.
\epf

\begin{cor}
Let $Z$ be a projective coherent space. The canonical scalar
multiplication on the nondegeneration space $[Z]$ is associative.
\end{cor}

\begin{thm}\label{t.class}
Let $Z$ be a coherent space and $A:Z\to Z$ be a coherent map with
adjoint $A^*$. Then the class map $[A]:[Z]\to[Z]$ defined by
\[
[A][z]:=[Az]\ \Forall z\in Z,
\]
is a well-defined and coherent map with the unique adjoint
$[A]^*=[A^*]$.
\end{thm}

\bepf
Let $z,z'\in Z$ with $[z]=[z']$. Using coherence of $A$, we have
\[
K(Az,z'')=K(z,A^*z'')=K(z',A^*z'')=K(Az',z''),
\]
for all $z''\in Z$. Thus, $[Az]=[Az']$ and hence $[A]:[Z]\to[Z]$ is
well-defined. Then, using coherence of $A$ and applying the definition of
the class map for the coherent maps $A$ and $A^*$, we get
\[
\bary{lll}
K([A][z],[z''])&=&K([Az],[z''])=K(Az,z'')=K(z,A^*z'')\\
&=&K([z],[A^*z''])=K([z],[A^*][z''])

\eary
\]
for all $z,z''\in Z$. This guarantees that the class map $[A]$ is a
coherent map with the unique adjoint $[A]^*=[A^*]$.
\epf

\begin{thm}
Let $Z$ be a coherent space. Then $[PZ]\cong P[Z]$, using a canonical
identification. In particular,

(i) if $Z$ is projective then we have $[PZ]\cong [Z]$.

(ii) if $Z$ is nondegenerate then $PZ$ is nondegenerate.
\end{thm}
\bepf
The canonical map $\rho:[PZ]\to P[Z]$ given by
$[(\lambda,z)]\to (\lambda,[z])$ is well-defined.
It is also straightforward to check that $\rho$ is a bijection.
Let $[(\lambda,z)],[(\lambda',z')]\in [PZ]$. Then, we have
\[
\bary{lll}
K_{\rm pe}(\rho[(\lambda,z)],\rho[(\lambda',z')])
&=&K_{\rm pe}((\lambda,[z]),(\lambda',[z']))
=\ol{\lambda}K([z],[z'])\lambda'\\
&=&\ol{\lambda}K(z,z')\lambda'=K_{\rm pe}((\lambda,z),(\lambda',z'))
=K_{\rm pe}([(\lambda,z)],[(\lambda',z')]),
\eary
\]
implying that $\rho:[PZ]\to P[Z]$ is an isomorphism of coherent spaces.
If $Z$ is projective then $PZ\cong Z$. Thus, we get $[PZ]\cong [Z]$.
If $Z$ is nondegenerate then $[Z]\cong Z$. Hence, we have
$[PZ]\cong P[Z]\cong PZ$, which implies that $PZ$ is nondegenerate as
well.
\epf

\section{Classical theory of functions of positive type}
\label{s.cohfuncpostype}

This section is independent of the remainder. It provides, in the
present physics-oriented terminology (cf. the introduction to
Section \ref{s.coh}) and with full proofs, a self-contained synopsis
(and sometimes slight generalization) of a number of classical results
from the literature about functions of positive type and reproducing
kernel Hilbert spaces.
In particular, the Moore--Aronszajn Theorem \ref{t.Moore} provides the
existence of the quantum space of a coherent space, hence is of
fundamental importance. However, on first reading, this theorem
can be taken for granted, and the study of the remainder of the section
can be postponed until the material is needed in later papers on
coherent spaces.

\subsection{The Moore--Aronszajn theorem}\label{ss.Moore}

This section discusses how to reconstruct a Hilbert space from a
spanning set of vectors whose inner product is known, and the
properties that must be satisfied for arbitrarily assigned formal inner
products to produce a Hilbert space.

The following theorem is due to \sca{Aronszajn} \cite{Aro} (1943), who
attributed\footnote{
\sca{Aronszajn} \cite[Th\'eor\`eme 2]{Aro0} states the theorem
and gives a detailed proof (in French), but his later English paper
\cite{Aro} states the theorem on p.344 and attributes it to Moore.
He cites \sca{Moore} \cite{Moo} (and a very short notice from 1916)
on p.338, but the theorem does not seem to be in one of these
references.
(Moore discusses in Chapter III functions of positive type under the
name {\em positive Hermitian matrices} -- cf. the statement at the top
of p.182 -- but does not construct a Hilbert space from them.)
\sca{Faraut \& Kor\'anyi} \cite[p. 170]{FarK} ascribes the theorem to
\sca{Bergman} \cite{Ber} (1933), but the theorem does not seem to be
there either. \sca{Kolmogorov} \cite[Lemma 2]{Kol} (1941) contains
the result for the special case where $Z$ is countable.
} 
it to Moore (1935).

\begin{thm}\label{t.Moore} (Moore, Aronszajn)~\\
Let $K:Z\times Z\to \Cz$ be of positive type. Then there is a unique
Hilbert space $\ol\Qz $ of complex-valued functions on $Z$ with
the Hermitian inner product $\<\cdot,\cdot\>$ (antilinear in
the first component) such that the following properties hold.

(i) $\ol\Qz$ contains the functions $q_z:Z\to \Cz$ defined for $z\in Z$
by
\lbeq{cp4}
q_z(x):=K(x,z)=\ol{K(z,x)}.
\eeq
(ii) The space $\Qz$ of finite linear combinations of the $q_z$ is
dense in $\ol \Qz $.

(iii) The following relations hold:
\lbeq{cp5}
\<q_z,q_{x}\>=K(z,x),
\eeq
\lbeq{cp6}
\psi(z)=\<q_z,\psi\> ~~\mbox{ for all } \psi\in\ol\Qz.
\eeq
(iv) For each $z\in Z$, the linear functional $\iota_z$
defined by
\lbeq{e.iotaz}
\iota_z\psi:=\psi(z)
\eeq
is continuous.
\end{thm}

\bepf
The vector space $\Qz$ spanned by the $q_z$ consists of all linear
combinations
\lbeq{cp7}
\wh f:=\sum_z f(z)q_{z}
\eeq
with $f$ in the space $\Fz$ of all maps $f:Z\to \Cz$ for which all
but finitely many values $f(z)$ vanish. Thus the sum is finite,
and by \gzit{cp4}, function values can be calculated by
\lbeq{cp8}
\wh f(x)=\sum_z f(z)q_{z}(x)=\sum_z K(x,z)f(z).
\eeq
Since it might be possible that a function $\psi\in\Qz $ can be
written in several ways in the form \gzit{cp7}, the definition
of an inner product on $\Qz $ requires some care. The mapping
defined on $\Fz \times \Fz$ by
\lbeq{cp10}
(g,f):=\sum_{x} \ol{g(x)}\wh f(x)=\sum_{x,z} \ol {g(x)} K(x,z)f(z)
\eeq
is a Hermitian form since
\[
\ol{(g,f)}=\sum_{x,z} g(x)\ol{K(x,z)}\ol{f(z)}
=\sum_{x,z} \ol{f(z)}K(z,x)g(x)=(f,g).
\]

Now
\lbeq{e.gf2}
(g,f)=\ol{(f,g)}=\sum_{x}f(x)\ol{\wh g(x)}=\sum_{z}\ol{\wh g(z)}f(z).
\eeq
If $\wh g=\wh u$ and $\wh f=\wh v$ then
\[
\bary{lll}
(g,f)&=&\D\sum_{z}\ol{\wh g(z)}f(z)=\sum_{z}\ol{\wh u(z)}f(z)
=(u,f)\\[5mm]
&=&\D\sum_{x} \ol{u(x)}\wh f(x)=\sum_{x} \ol{u(x)}\wh v(x)=(u,v).
\eary
\]
Therefore $(g,f)$ depends only on the functions $\wh g$ and $\wh f$.
Thus
\[
\<\psi,\psi'\>:=(g,f)~~\mbox{ if } \psi=\wh g,~\psi'=\wh f
\]
defines a Hermitian form $\<\cdot,\cdot\>$ on $\Qz$ satisfying
\lbeq{e.ip}
\<\wh g,\wh f\>=(g,f).
\eeq
The function $g_z\in \Fz$ defined (for arbitrary but fixed $z\in Z$)
by $g_z(x)=1$ if $x=z$ and $g_z(x)=0$ otherwise, satisfies
\lbeq{e.gz}
\wh g_z=q_z
\eeq
by \gzit{cp7}, hence by \gzit{e.gf2},
\[
\<q_z,\wh f\>=\<\wh g_z,\wh f\>=(g,f)
=\sum_{x}\ol{g_z(x)}\wh f(x)=\wh f(z).
\]
Since by definition of $\Qz$, any $\psi\in \Qz $ can be
written as $\psi=\wh f$, we conclude \gzit{cp6}. Specialization to
$\psi=q_x$ and using \gzit{cp4} (with $z$ and $x$ interchanged)
yields \gzit{cp5}.

Since $K$ is of positive type, $(f,f)\ge 0$ for all $f\in\Fz$.
Thus the form is positive semidefinite on $\Fz$. In particular, the
Cauchy--Schwarz inequality $|(f,f')|^2\le (f,f)(f',f')$ holds.
It implies that $(f,f)=0$ only if $(f,f')=0=(f',f)$ for all $f'$,
and \gzit{cp7} then shows that $\wh f(z)=0$ for all $z$. Hence
$\wh f=0$. Therefore the Hermitian form $\<\cdot,\cdot\>$ is positive
definite, hence defines a Hermitian inner product on $\Qz$.
Thus $\Qz$ is a Euclidean space. The completion with respect to the norm
\[
\|\psi\|:=\sqrt{\<\psi,\psi\>}
\]
(which can be done constructively using Theorem \ref{t.Hilbert})
gives the desired Hilbert space, and a limiting argument shows that
\gzit{cp6} holds in general:
If $\psi\in\ol\Qz$, there is a net of $\psi_j\in\Qz$ converging
to $\psi$ in the norm, and
\[
|\<q_z,\psi\>-\psi_j(z)|=|\<q_z,\psi\>- \<q_z,\psi_j\>|
=|\<q_z,\psi-\psi_j\>|\le \|q_z\|\,\|\psi-\psi_j\|\to 0,
\]
hence $\psi(z)=\D\lim_j \psi_j(z)\to \<q_z,\psi\>$.

(iv) Since $\iota_z\psi=\psi(z)=\<q_z,\psi\>$, we have
$\|\iota_z\|=\|q_z\|$. Thus $\iota_z$ is bounded and hence continuous.

The uniqueness of $\ol\Qz$ is clear from the construction.
\epf

\subsection{Reproducing kernel Hilbert spaces and Mercer's theorem}

\begin{prop} \label{p.Q-CountOrthoNormBasis}
Let $\psi_\alpha$ ($\alpha\in I$) be an orthonormal basis for $\ol\Qz$.
Then
\lbeq{cp12}
K(z,w)=\sum_{\alpha\in I}\psi_\alpha(z)\ol{\psi_\alpha(w)}.
\eeq
\end{prop}

\bepf
By the polarized version of the Parseval identity,
Theorem 5.27 of \sca{Folland} \cite{Foll.R},
we have
\[
q_w=\sum_{\alpha\in I}\<\psi_\alpha,q_w\>\psi_\alpha
=\sum_{\alpha\in I}\ol{\psi_\alpha(w)}\psi_\alpha
\]
for all $w\in Z$. Hence for all $z,w\in Z$,
\[
K(z,w)=\<q_z,q_w\>
=\Big\<q_z,\sum_{\alpha\in I}\ol{\psi_\alpha(w)}\psi_\alpha\Big\>
=\sum_{\alpha\in I}\<q_z,\psi_\alpha\>\ol{\psi_\alpha(w)}
=\sum_{\alpha\in I}\psi_\alpha(z)\ol{\psi_\alpha(w)},
\]
which implies (\ref{cp12}).
\epf

A \bfi{reproducing kernel Hilbert space} is a Hilbert space $\Kz$ of
functions on a set $Z$ together with a \bfi{reproducing kernel}
$K:Z\times Z\to \Cz$ such that the functions\footnote{
Slightly more generally, a reproducing kernel Hilbert space may be
defined as a Hilbert space $\Kz$ of functions on a set $Z$ with an
involution $\,\ol{\phantom{g}}\,$ together with a reproducing kernel
$K:Z\times Z\to \Cz$ such that the functions $k_z$ ($z\in Z$) defined by
$k_z(x):=K(\ol x,z)$ span a space dense in $\Kz$ and satisfy
$\psi(z)=k_{\ol z}^*\psi$ for all $\psi\in\Kz,~z\in Z$.
If we define, with an arbitrary choice of an involution $\sim$ on $Z$,
for $\psi\in\ol\Qz$ the function $\wt\psi:Z\to\Cz$ by
$\wt\psi(z):=\psi(\wt z)$, \gzit{cp6} says that
$\wt\Qz:=\{\psi \mid \wt\psi\in\ol \Qz\}$ is a reproducing kernel
Hilbert space with reproducing kernel $K$ and $k_z=\wt q_z$.
This generalization (which just amounts to a relabeling of the
arguments of the functions $k_z$) is useful when considering sets $Z$
with the structure of a complex manifold, and wants the functions
$k_z$ to be analytic rather than antianalytic.
} 
$k_z$ ($z\in Z$) defined by
\lbeq{e.qzz}
k_z(x):=K(x,z)
\eeq
span a space dense in $\Kz$ and satisfy
\lbeq{e.psiq}
\psi(z)=k_{z}^*\psi \Forall \psi\in\Kz,~z\in Z.
\eeq
If we define for $\psi\in\ol\Qz$ the function $\wt\psi:Z\to\Cz$ by
\[
\wt\psi(z):=\psi(\wt z),
\]
\gzit{cp6} says that $\wt\Qz:=\{\psi \mid \wt\psi\in\ol \Qz\}$ is a
reproducing kernel Hilbert space with reproducing kernel $K$ and
$k_z=\wt q_z$.

Proposition \ref{p.Q-CountOrthoNormBasis} is related to Mercer's theorem
(\sca{Mercer} \cite{Mer}), which represents certain reproducing kernels
by an infinite sum of the form
\[
K(z,w)=\sum_{\alpha\in I}\lambda_\alpha\phi_\alpha(z)\ol{\phi_\alpha(w)}
\]
with positive real numbers $\lambda_\alpha$ and functions $\phi_\alpha$
satisfying additional properties. Precise statements of Mercer's
theorem and its generalizations (e.g., \sca{Ferreira \& Menegatto}
\cite{FerM}) require additional structure on $Z$ and $K$ concerning
measurability and continuity, hence are not valid in the generality
discussed here. We therefore refrain here from giving details, and refer
to a future paper in this series for the discussion of measure theoretic
properties of coherent states and associated overcompleteness relations.

\subsection{Theorems by Bochner and Kre\u\i n}

\sca{Bochner} \cite[Satz 4]{Boc} proved the following optimality
result for $q_x$.

\begin{thm} (Bochner)\\
Let $K:Z\times Z\to\Cz$ be of positive type, and let $\Qz$ be the space
constructed in the Moore--Aronszejn theorem (Theorem \ref{t.Moore}).
If $x\in Z$ satisfies $K(x,x)\ne 0$ then
\[
\min\{\psi^*\psi\mid \psi\in\Qz,~\psi(x)=\alpha\}
= \frac{|\alpha|^2}{K(x,x)}.
\]
The minimum is attained just for $\psi=\D \frac{\alpha}{K(x,x)}q_x$.
In particular, if $\alpha=K(x,x)$, the minimum is attained precisely
at $q_x$.
\end{thm}

\bepf
This is trivial for $\alpha=0$. For $\alpha\ne0$ we may rescale the
assertion; thus it is enough to prove the case $\alpha=K(x,x)$.
In this case
\[
\bary{lll}
\psi^*\psi&=&\<\psi-q_x,\psi-q_x\>+2\re\<q_x,\psi\>-\<q_x,q_x\>\\
&=&\<\psi-q_x,\psi-q_x\>+2\re \psi(x)-K(x,x)
=\|\psi-q_x\|^2+K(x,x)\ge K(x,x)=\alpha,
\eary
\]
with equality iff $\psi-q_x=0$.
\epf

Our next result, a variant of \sca{Kre\u\i n} \cite{Kre1}, characterizes
which functions $\psi\in\Qz^\*$ belong already to the Hilbert space
$\ol \Qz$.

\begin{thm} \label{t.Krein} (Kre\u\i n)\\
Let $K:Z\times Z\to\Cz$ be of positive type and $\psi:Z\to\Cz$.
Define the function
$K_\eps:Z\times Z\to\Cz$ by
\[
K_\eps(z,z'):=K(z,z')-\eps\psi(z)\ol{\psi(z')}.
\]
(i) If $\psi\in\ol\Qz$ and $0<\eps\le \|\psi\|^{-2}$ then $K_\eps$ is of
positive type.

(ii) If $K_\eps$ is of positive type for some $\eps>0$ then
$\psi\in\ol\Qz$.
\end{thm}

\bepf
(i) Hermiticity is obvious. To show that $K_\eps$ is of positive type
we need to show for any finite sequence of complex numbers $u_k$ and
points $z_k\in Z$ the nonnegativity of the sum
\[
\sigma:=\sum_{j,k} \ol u_jK_\eps(z_j,z_k)u_k
=\sum_{j,k} \ol u_j\<q_{z_j},q_{z_k}\>u_k
-\eps \sum_{j,k} \ol u_j\psi(z_j)\ol{\psi(z_k)}u_k,
\]
where we used \gzit{cp5}. Writing
\[
q:=\sum_k q_{z_k}u_k,
\]
we find that
\[
\<\psi,q\>=\sum_k \<\psi,q_{z_k}\>u_k
=\sum_k \ol{\<q_{z_k},\psi\>}u_k
=\sum_k \ol{\psi(z_k)}u_k,
\]
hence
\[
\sigma=\|q\|^2-\eps|\<\psi,q\>|^2
\ge \|q\|^2-\eps\|\psi\|^2\|q\|^2\ge 0.
\]
(ii) In this case, with $\Fz$ and $\wh f$ as in the proof of the
Moore--Aronszejn theorem (Theorem \ref{t.Moore}), we consider the
antilinear mapping $\Psi:\Fz\to\Cz$ defined by
\[
\Psi(f):=\sum_{z\in Z}\ol{f(z)}\psi(z).
\]
Since $K_\eps$ is of positive type, we have
\[
\bary{lll}
0&\le&\D\sum_{z,z'}\ol{f(z)}K_\eps(z,z')f(z')
=\D\sum_{z,z'}\ol{f(z)}K(z,z')f(z')
-\eps\sum_{z,z'}\ol{f(z)}\psi(z)\ol{\psi(z')}f(z')\\[5mm]
&=&(f,f)-\eps|\Psi(f)|^2
=\|\wh f\|^2-\eps|\Psi(f)|^2
\eary
\]
by definition of $K_\eps$, \gzit{cp10}, \gzit{e.ip}, and the definition
of $\Psi$. Therefore
\[
|\Psi(f)|\le \eps^{-1/2}\|\wh f\|.
\]
In particular, $\wh f=0$ implies $\Psi(f)=0$. Therefore $\Psi$ defines
a unique antilinear mapping $\psi':\Qz\to\Cz$ with
$\psi'(\wh f)=\Psi(f)$ for all $f\in\Fz$. By the above,
$|(\psi'\wh f)|\le \eps^{-1/2}\|\wh f\|$. Thus
$\psi'$ is bounded. By Theorem \ref{t.Hilbert}, $\wh\Psi$ belongs to
\[
\wh \Psi(\phi)=\<\phi,\psi'\> \Forall \psi\in\ol \Qz.
\]
Since by \gzit{e.gz} and \gzit{cp6},
\lbeq{e.bad}
\psi(z)=\Psi(g_z)=\wh\Psi(\wh g_z)=\wh\Psi(q_z)=\<q_z,\psi'\>=\psi'(z)
\eeq
for all $z\in Z$, we conclude that $\psi=\psi'\in\ol\Qz$.
\epf

\subsection{Theorems by Schoenberg and Menger}\label{ss.SM}

In this subsection we prove the promised converse of Propositions
\ref{p.condP0}--\ref{p.condP}.

\begin{thm}\label{t.condpos} (\sca{Schoenberg} \cite[p.49]{Scho})\\
If $F$ is conditionally positive then the function $P_a$, defined for
any $a\in Z$ by
\lbeq{e.SF}
P_a(z,z'):= F(z,z')-F(z,a)-F(a,z')+F(a,a),
\eeq
is of positive type. Conversely, if a map $F:Z\times Z\to\Cz$ is such
that if $P_a$ is of positive type for {\em some} $a\in Z$ then  $F$ is
conditionally positive.
\end{thm}

\bepf
Let $G,\wt G$ be the Gram matrices of $z_1,\ldots,z_n$ computed with
$F$ and $P_a$, respectively. Then
\[
\wt G=G-g\1^*-\1 g^*+\gamma \1\1^*,
\]
where $\1$ is the all-one column vector, $g$ the column vector with
components $g_j:=F(z_j,a)$, and $\gamma:=F(a,a)$. The Gram matrix of
$z_1,\ldots,z_n,a$ computed with $F$ is therefore
\[
G':=\pmatrix{G & g\cr g^* & \gamma}.
\]
Now $v\in\Cz^{n+1}$ satisfies $\D\sum_j v_j=0$ iff, for some
$u\in\Cz^n$,
\[
v=\pmatrix{u \cr -s},~~~s=\1^*u,
\]
and then
\[
\bary{lll}
v^*G'v
&=&\pmatrix{u \cr -s}^*\pmatrix{G & g\cr g^*& \gamma}\pmatrix{u \cr -s}
=u^*Gu-u^*gs-\ol s g^*u+\gamma \ol s s\\[4mm]
&=&u^*(G-g\1^*-\1 g^*+\gamma \1\1^*)u=u^*\wt G u.
\eary
\]
This shows that $F$ is conditionally positive iff all $\wt G$ are
positive semidefinite, i.e., iff $P_a$ is of positive type for some $a$
and hence for all $a$.
\epf

\begin{thm}\label{t.Menger}
A map $F:Z\times Z\to \Cz$ is conditionally positive iff there is an
embedding $z\to q_z$ of $Z$ into a Euclidean space $\Hz$ such that
\lbeq{e.Ff}
F(z,z')=\ol{f(z)}+f(z')+q_z^*q_{z'}
\eeq
holds for some $f:Z\to\Cz$.
\end{thm}

\bepf
(i) Suppose that $F$ is conditionally positive. Fix $a$ and define
$P_a$ by \gzit{e.SF}.
By Theorem \ref{t.condpos}, $P_a$ is of positive type. Hence the
Moore--Aronszejn theorem (Theorem \ref{t.Moore}) gives an embedding
$z\to q_z$ into a Hilbert space such that
\lbeq{e.Paqq}
P_a(z,z')=q_z^*q_{z'},
\eeq
applying \gzit{cp5} of the theorem to $P_a$ in place of $K$.
The definition of $P_a$ then implies
\[
F(z,z')-F(z,a)-F(a,z')+F(a,a)=q_z^*q_{z'}.
\]
Putting $z=z'=a$ gives $q_a^*q_a=0$, hence $q_a=0$. One now easily
verifies that
\[
D(z,z'):=F(z,z')-q_z^*q_{z'}
\]
satisfies $\ol{D(z,z')}=D(z',z)$ and
\[
D(z,z')-D(z,a)-D(a,z')+D(a,a)=0.
\]
This implies that $D(z,z')=\ol{f(z)}+f(z')$ with
\[
f(z):=D(a,z)-\half D(a,a).
\]
Therefore \gzit{e.Ff} holds.

(ii) Conversely, if \gzit{e.Ff} holds then \gzit{e.Paqq} and \gzit{e.SF}
imply that $P_a(z,z')=\<q_z-q_a,q_{z'}-q_a\>$, hence $P_a$ is of
positive type.
By Schoenberg's Theorem \ref{t.condpos}, $F$ is conditionally positive.
\epf

The following converse of Proposition \ref{p.condP} is related to
results by \sca{Menger} \cite{Men} in the context of characterizing
metric spaces embeddable into a finite-dimensional real vector space.

\begin{cor}\label{c.cp}
A map $F:Z\times Z\to \Cz$ satisfying $F(z,z')=F(z',z)$ for $z,z'\in Z$
is conditionally positive iff there is an embedding $z\to q_z$ of $Z$
into a real Euclidean space such that
\lbeq{e.Ffq}
F(z,z')=g(z)+g(z')-\|q_z-q_{z'}\|^2
\eeq
holds for some $g:Z\to\Rz$.
\end{cor}

\bepf
If there is such an embedding then $F$ is conditionally positive by
Proposition \ref{p.condP}. Conversely, suppose that $F$ is conditionally
positive. Then $\half F$ is also conditionally positive.
By Theorem \ref{t.Menger}, there is an embedding $z\to q_z$ of $Z$
into a complex Euclidean space $\Hz$ such that
\[
\half F(z,z')=\ol{f(z)}+f(z')+q_z^*q_{z'}
\]
holds for some $f:Z\to\Cz$. Since we assumed $F(z,z')=F(z',z)$ for
$z,z'\in Z$, $f(z)$ is real. Thus the inner products
$K(z,z')=q_z^*q_{z'}$ are also real, and the $q_z$ span a real
Euclidean space. Substitution of $g(z)=2f(z)+\|q_z\|^2$ now shows that
\gzit{e.Ffq} holds.
\epf

\subsection{The Berezin--Wallach set}\label{ss.BWset}

Many coherent products of interest have the exponential form discussed
in the following theorem. It is due to \sca{Schoenberg} \cite{Scho42} in
the case where $F$ takes only finite values and is zero on the diagonal,
to \sca{Herz} \cite[Proposition 6]{Herz} in the case where $F$ takes
only finite values, and to \sca{Horn} \cite{Hor} in the general case.
The present proof is much shorter than Horn's.

To be able to formulate the results, we put
\[
e^{-\infty}:=0
\]
and call a function $F:Z\times Z\to \Cz\cup\{-\infty\}$
\bfi{conditionally positive} if either
(i) there is an equivalence relation $\equiv$ on $Z$ such that $F$ is
conditionally positive on each equivalence class, and $F(z,z')=-\infty$
whenever $z\not\equiv z'$, or
(ii) $F$ takes only infinite values.
This reduces to the original definition if the value $-\infty$ is not
attained, which holds iff there is only one equivalence classe.

\begin{thm} \label{t.divis}~\\
(i) If $F:Z\times Z\to \Cz\cup\{-\infty\}$ is conditionally positive
then, for all $\beta>0$,
\lbeq{e.div}
K(z,z'):=e^{\beta F(z,z')}
\eeq
is of positive type.

(ii) Let $F:Z\times Z\to \Cz\cup\{-\infty\}$.
If there is a sequence of positive numbers $\beta_k$ converging to $0$
such that
\[
K_k(z,z'):=e^{\beta_k F(z,z')}
\]
is of positive type for all $k$ then $F$ is conditionally positive.
\end{thm}

\bepf
(i) If $F$ takes only finite values then Theorem \ref{t.condpos} shows
that (for any $z_0\in Z$), the function $\wt F$ defined by
\[
\wt F(z,z'):= \beta(F(z,z')-F(z,z_0)-F(z_0,z')+F(z_0,z_0))
\]
is of positive type. Theorem \ref{t.powers} therefore implies that
$K(z,z'):= e^{\wt F(z,z')}$ defines a function $K$ of positive type.
Rescaling this by Proposition \ref{p.cons}(iv), we see that
\gzit{e.div} is of positive type, too.
If $F$ takes infinite values only, $K$ is identically zero and hence of
positive type. If $F$ takes finite and infinite values, the previous
argument may be applied to the restriction of $K$ to each equivalence
class, and shows that this restriction is of positive type. Then
Proposition \ref{p.cons}(vi) implies that $K$ itself is of positive
type.

(ii) We may assume w.l.o.g. that $Z$ cannot be decomposed
as in Proposition \ref{p.cons}(vi).

\sca{Case 1:} $K(z,z')\ne 0$ for all $z,z'\in Z$.
We fix $a\in Z$ and use Proposition \ref{p.cons}(iv) to rescale
$K:=e^{\beta F}$ (consistently for all $\beta$) such that all
$K(a,z)=1$, hence all $F(a,z)$ vanish. Theorem \ref{t.condpos}, applied
with $K_k$ in place of $F$, implies that the map $P_a:Z\times Z\to\Cz$
defined by
\[
P_a(z,z')=K_k(z,z')-K_k(z,a)-K_k(a,z')+K_k(a,a)=K_k(z,z')-1
\]
is of positive type. Therefore the functions $F_k$ defined by
\[
F_k(z,z'):=\frac{K_k(z,z')-1}{\beta_k}
=\frac{e^{\beta_kF(z,z')}-1}{\beta_k}
= F(z,z')+\beta_kF(z,z')^2+O(\beta_k)^2
\]
are also of positive type. Since $\beta_k\to 0$, $F_k(z,z')\to F(z,z')$
for $k\to\infty$. By Proposition \ref{p.cons}(vii), $F$ is of positive
type. Undoing the scaling and using Theorem \ref{t.condpos} now proves
that $F$ is conditionally positive.

\sca{Case 2:} $K(z,z)=0$ for some $z\in Z$. Then the positivity of
the Gram matrix \gzit{e.gram} for $n=2$ implies that $K(z,z')=K(z',z)=0$
for all $z'\in Z$, and the indecomposability assumed at the beginning
of (ii) implies that $Z=\{z\}$ and $K$ is identically zero. Thus $F$
takes only infinite values and is therefore conditionally positive.

\sca{Case 3:} $K(z,z)\ne 0$ for all $z\in Z$ but $K(x,y)=0$ for some
$x,y\in Z$. By Proposition \ref{p.cons}(iv) we may normalize $K$ such
that $K(z,z)=1$ for all $z\in Z$. The Gram matrix
\[
G=\pmatrix{K(x,x) & K(x,y) & K(x,z)\cr
           K(y,x) & K(y,y) & K(y,z)\cr
           K(z,x) & K(z,y) & K(z,z)}
=\pmatrix{1 & K(x,y) & K(x,z)\cr
           K(y,x) & 1 & K(y,z)\cr
           K(z,x) & K(z,y) & 1}
\]
of $x,y,z\in Z$ is Hermitian and poitive semidefinite, hence its
determinant is nonnegative,
\[
0\le 1-|K_k(x,z)|^2-|K_k(y,z)|^2
=1-|K(x,z)|^{2\beta_k}-|K(y,z)|^{2\beta_k}.
\]
Unless at least one of $K(x,z)$ or $K(y,z)$ vanishes, the two negative
terms tend for $k\to\infty$ both to $1$, hence the right hand side
converges to $-1$. This holds for any $z$, whence $Z$
can be split into two subsets $X$ and $Y$ such that $K(x,z)=0$ for
$z\in Y$ and $K(y,z)=0$ for $z\in X$. By Hermiticity, $K(z,x)=0$ for
$z\in Y$ and $K(z,y)=0$ for $z\in X$. Repeating the argument for all
zeros constructible this way shows that $Z$ decomposes as in
Proposition \ref{p.cons}(iii), contradiction.
\epf

If we want to discuss a possible generalization of Theorem
\ref{t.powers} to other exponents we need to assume that the power
exists, which suggests to assume for $K(z,z')$ an exponential form.

The \bfi{Berezin--Wallach set} of a mapping
$F:Z\times Z\to \Cz\cup\{-\infty\}$ is the set $W(F)$ of nonnegative
real numbers $\beta$ for which
 \lbeq{e.BW}
K(z,z'):=e^{\beta F(z,z')}
\eeq
is of positive type. The \bfi{Berezin--Wallach set} of a coherent space
is the set $W(F)$ where
\[
F(z,z'):=\log K(z,z'),
\]
using the principal value of the logarithm and $\log 0=-\infty$.
(Thus the Berezin--Wallach set of a coherent space always contains $1$.)

This set was introduced by \sca{Wallach} \cite{Wal} in the context of
representations of Lie groups.  But already earlier, \sca{Berezin}
\cite{Berez} computed the Berezin--Wallach set for the case when
$F$ is the K\"ahler potential of a Siegel domain. Indeed, in many cases
of interest, $Z$ is a so-called K\"ahler manifold and $F$ the associated
K\"ahler potential; see, e.g.,
\sca{Zhang} et al. \cite[Section VI]{ZhaFG}.
For the Berezin--Wallach sets corrsponding to Hermitian symmetric
spaces see, e.g., \sca{Faraut \& Kor\'anyi} \cite[Section XIII.2]{FarK}.

\begin{thm}\label{t.BW}~\\
(i) The Berezin--Wallach set $W(F)$ is a closed set containing $0$.

(ii) $W(F)$ contains with $\beta$ and $\beta'$ their sum and hence all
linear combinations with nonnegative integral coefficients.

(iii) If $W(F)$ contains an open set it contains all sufficiently
large positive real numbers.

(iv) If $F$ is conditionally positive then $W(F)$ contains all
nonnegative real numbers.

(v) If $0$ is a limit point of $W(F)$ then $F$ is conditionally
positive.
\end{thm}

\bepf
(i)--(iv) follow easily from Proposition \ref{p.cons2}(ii).
(v) follows from Theorem \ref{t.divis}.
\epf

In the most interesting cases, the Berezin--Wallach set is of the form
$\alpha\Nz_0\cup[\beta,\infty]$ or $\alpha\Nz_0$, where
$\alpha,\beta>0$ and $\Nz_0$ denotes the set of nonnegative integers.
In general, the Berezin--Wallach set may have a very complicated
structure, already for $Z$ with three elements only.
\sca{FitzGerald \& Horn} \cite{FitH} show that the Berezin--Wallach set
of every finite coherent space $Z$ with real, nonnegative coherent
product contains the interval $[|Z|-2,\infty[$.

\bigskip

\end{document}